\documentclass[11pt]{article}
\usepackage[letterpaper,margin=0.9in]{geometry}
\usepackage[T1]{fontenc}
\usepackage[utf8]{inputenc}
\usepackage{newtxtext}
\usepackage{url}
\usepackage[font=small,labelfont=bf]{caption}
\usepackage{graphicx}
\usepackage{amsmath}
\usepackage{amsthm}
\usepackage{aliascnt}
\usepackage{newtxmath}
\usepackage{mathtools}
\usepackage{booktabs}
\usepackage{tabularx}
\usepackage{array}
\usepackage{enumitem}
\usepackage{xcolor}
\usepackage{microtype}
\usepackage{authblk}
\usepackage{natbib}
\usepackage{placeins}
\usepackage[hidelinks]{hyperref}
\usepackage[nameinlink,noabbrev]{cleveref}

\hypersetup{
  pdftitle={Dynamic Welfare-Maximizing Pooled Testing},
  pdfauthor={Nicholas Lopez, Francisco Marmolejo-Cossio, Jose Roberto Tello Ayala, David C. Parkes}
}

\newtheorem{theorem}{Theorem}[section]
\newtheorem*{harrisfkg}{Harris--FKG inequality}
\newaliascnt{lemma}{theorem}
\newtheorem{lemma}[lemma]{Lemma}
\aliascntresetthe{lemma}
\newaliascnt{proposition}{theorem}
\newtheorem{proposition}[proposition]{Proposition}
\aliascntresetthe{proposition}
\newaliascnt{corollary}{theorem}
\newtheorem{corollary}[corollary]{Corollary}
\aliascntresetthe{corollary}
\newaliascnt{conjecture}{theorem}

\aliascntresetthe{conjecture}
\theoremstyle{definition}
\newaliascnt{definition}{theorem}
\newtheorem{definition}[definition]{Definition}
\aliascntresetthe{definition}
\newaliascnt{example}{theorem}
\newtheorem{example}[example]{Example}
\aliascntresetthe{example}
\newaliascnt{hypothesis}{theorem}
\newtheorem{hypothesis}[hypothesis]{Hypothesis}
\aliascntresetthe{hypothesis}
\theoremstyle{remark}
\newaliascnt{remark}{theorem}
\newtheorem{remark}[remark]{Remark}
\aliascntresetthe{remark}
\newaliascnt{openproblem}{theorem}

\aliascntresetthe{openproblem}

\setlist{nosep}

\newcommand{\E}{\mathbb{E}}
\newcommand{\Prob}{\mathbb{P}}
\newcommand{\1}{\mathbf{1}}
\newcommand{\OPTdyn}{\operatorname{OPT}_{\mathrm{dyn}}}
\newcommand{\OPTstar}{\operatorname{OPT}^{*}}
\newcommand{\OPTno}{\operatorname{OPT}}
\newcommand{\Ano}{\mathcal A_{\mathrm{no}}}
\newcommand{\Agap}{\mathcal A}
\newcommand{\Greedy}{\textnormal{\textsc{Exact-Joint Greedy}}}
\newcommand{\StaticGreedy}{\textnormal{\textsc{Static Greedy}}}
\newcommand{\MPGreedy}{\textnormal{\textsc{Marginal-Product Greedy}}}

\title{
Dynamic Welfare-Maximizing Pooled Testing\thanks{
Code and data accompanying this paper will be released at
\url{https://github.com/nrlopez03/pooled-testing}.
}
}

\author[1]{Edwin Lock}
\author[2]{Nicholas Lopez}
\author[3]{Francisco Marmolejo-Cossío}
\author[2]{Jose Roberto Tello Ayala}
\author[2]{David C. Parkes}
\affil[1]{King's College London}
\affil[2]{Harvard University}
\affil[3]{Boston College}
\affil[ ]{\texttt{edwin.lock@kcl.ac.uk}, \texttt{nicholaslopez@alumni.harvard.edu}, \texttt{jtelloayala@g.harvard.edu}, \texttt{marmolf@bc.edu}, \texttt{parkes@eecs.harvard.edu}}

\date{}

\begin{document}

\maketitle

\begin{abstract}
Pooled testing uses one test to certify several agents as healthy when the pooled result is negative. We study a budget-constrained welfare problem in which agents have heterogeneous utilities and independent prior probabilities of being healthy.  Welfare is earned when an agent is certified healthy, and a dynamic policy may choose each pool after observing earlier test outcomes.  We ask how much such adaptation can improve over a static allocation that fixes all pools in advance.

Our main result proves that the optimal dynamic policy has value at most twice that of the optimal static overlapping allocation, for every population, test budget, and pool-size cap. The proof samples a path through the dynamic policy using an independent health profile, randomly thins the selected pools, and compares the resulting static allocation with the dynamic policy one agent at a time. A Boolean-cube argument proves the comparison on uniform subcubes, and a complementary-profile coupling lifts the result to arbitrary heterogeneous product priors. 

We also identify regimes in which adaptivity has no value, show that strict adaptive gains require re-pooling agents after positive tests, and give a three-agent instance in which adaptation is strictly beneficial. Exact-Joint Greedy obtains a $1/(e+1)$ fraction of optimal static overlapping welfare. The static non-overlapping Greedy algorithm of Finster et al. has the same guarantee, hence our factor-two theorem newly implies that each is a $2(e+1)$-approximation to the optimal dynamic policy. Finally, a reproducible exact small-instance study compares the dynamic and static benchmarks and greedy policies. The appendix records separate exploratory results for Gibbs-marginal and reinforcement-learning approaches at larger scales.
\end{abstract}

\section{Introduction}
\label{sec:introduction}

Public-health screening often operates with a fixed number of tests and a population too large for individual testing. Pooled testing stretches that budget by combining several samples in one test. A pool tests negative exactly when every person in it is healthy; a negative result therefore certifies, or \emph{clears}, all members of that pool.  In many applications the immediate objective is not to diagnose every infection, but to clear people for work, school, travel, or other activities.

We model a population of $n$ agents.  Agent $i$ is healthy with known prior probability $q_i$, and these health states are independent across agents.  Clearing agent $i$ produces utility $u_i\ge0$, which may reflect the social value of that agent's return to an activity.  Each test may contain at most $G$ agents, and at most $B$ tests may be used. An agent's utility is counted once, when the agent first appears in a negative pool. The objective is to maximize total expected clearance utility.

A \emph{static} allocation selects up to $B$ pools before observing any outcomes. We allow its pools to overlap, so the same agent may appear in more than one test. A \emph{dynamic} policy selects the first pool, observes whether it is positive or negative, and then chooses the next pool as a function of the observed history.  Let $J=([n],(u_i,q_i)_{i=1}^n,G)$ denote the complete instance, including its pool-size cap.  We write $\OPTstar(B,J)$ for the maximum expected welfare of a static overlapping allocation and $\OPTdyn(B,J)$ for the maximum expected welfare of a dynamic policy.  Since a dynamic policy can ignore outcomes and follow any fixed allocation, $\OPTdyn(B,J)\ge\OPTstar(B,J)$. Our central question is therefore:
\begin{quote}
How large can the ratio $\OPTdyn(B,J)/\OPTstar(B,J)$ be?
\end{quote}
We call this ratio the \emph{adaptivity gap}.  Our main theorem gives the universal bound
\[
  \OPTdyn(B,J)\le 2\OPTstar(B,J),
\]
for every collection of utilities and independent health probabilities, every budget $B$, and every pool-size cap $G$. A three-agent example in \cref{subsec:strict-example} shows that the ratio can be strictly larger than one, so observing outcomes can genuinely improve welfare.

The difficulty in obtaining a constant bound comes from two sources. First, a positive pooled test reveals that at least one member is infected without identifying which one, and later tests can exploit this information by re-testing some of those agents in different pools. Such overlap creates statistical dependence among their posterior health states. Second, a dynamic policy may choose a different sequence of pools for every possible test history. A comparison with a single fixed allocation must therefore control both the dependence created by overlapping tests and the policy’s history-dependent branching.

The proof converts an arbitrary dynamic policy into a randomized static allocation.  An independent health profile is used to select one sequence of pools that the policy would execute. Every agent occurrence in those pools is then retained independently with probability $1/2$, and the resulting pools are evaluated on a second, independent profile. We prove separately for each agent that this static construction clears the agent with at least half its dynamic clearance probability. The combinatorial argument first fixes some health coordinates and lets the others vary uniformly (a Boolean subcube) and then extends the result to heterogeneous product priors by conditioning two independent profiles coordinate by coordinate.
The complete argument appears in \cref{sec:factor-two}.

The paper also develops several structural and algorithmic consequences:
\begin{enumerate}[leftmargin=*]
\item After any observed history, the joint probability that a specified set of not-yet-cleared agents is entirely healthy is no greater than the product of their original health probabilities. This \emph{posterior-domination} property yields general welfare upper bounds and several regimes in which dynamic testing offers no advantage.
\item If the pools encountered along every possible execution are disjoint, conditioning future choices on earlier outcomes provides no benefit over an optimal static non-overlapping allocation. Thus every strict adaptive gain must involve re-pooling an agent after a positive test.
\item $\Greedy$ chooses, after each history, the feasible pool with the largest true conditional expected immediate welfare.  We prove that it obtains at least a $1/(e+1)$ fraction of $\OPTstar(B,J)$. The exact-oracle static non-overlapping Greedy algorithm of \citet{finster2026static} has the same benchmark guarantee.  By the factor-two theorem, each therefore obtains at least a $1/[2(e+1)]$ fraction of $\OPTdyn(B,J)$, or equivalently is a $2(e+1)$-approximation to the optimal dynamic policy.
\end{enumerate}

After formally defining the model, we give the strict-adaptivity example, develop the structural results and greedy guarantee, present the complete factor-two proof, and report exact small-instance experiments.  The appendices contain supporting proofs and the computational details needed to reproduce the experiments and implement the larger-scale empirical approaches.

\subsection{Related Work}
\label{subsec:related}

\paragraph{Classical pooled testing and adaptive group testing.}
Pooled testing was introduced by~\citet{largepop} to reduce the number of tests needed to determine individual infection statuses. A large subsequent literature studies both non-adaptive and adaptive full-diagnosis objectives. Early work by~\citet{Sobel_Groll} and~\citet{Hwang1976} developed dynamic programs for adaptive group testing, including extensions by~\citet{Nebenzahl1973} to heterogeneous infection probabilities; modern treatments survey many further variants~\citep{grouptest}. Those objectives usually minimize the expected number of tests needed for diagnosis, whereas our test budget is fixed and our objective is the welfare obtained from clearing healthy agents.

\paragraph{Constrained and overlapping pooled testing.}
Operational work considers noisy tests, overlapping pools, and deployment constraints~\citep{grouptest,poolisolate,poolappscov}. Budget-constrained designs also arise in reopening and screening problems~\citep{lock2021optimal,optriskgroup}. These formulations motivate the biological pool-size constraint in our model but generally optimize diagnostic or risk objectives rather than heterogeneous clearance welfare.

\paragraph{Welfare-maximizing pooled testing.}
The closest predecessor introduces static welfare-maximizing pooled testing, studies approximation algorithms, and reports a field deployment~\citep{welfaremax,finster2026static}. Finster et al.\ show that allowing overlapping pools in a static allocation can improve welfare by at most a factor of two over the best static non-overlapping allocation. Our result addresses a different question: we show that conditioning future pools on observed outcomes can improve welfare by at most a factor of two over the best static allocation that already permits overlap.
In addition, Finster et al.'s Greedy algorithm constructs a static non-overlapping allocation and, with an exact single-test oracle, obtains a $1/(e+1)$ fraction of the optimal static overlapping welfare~\citep[Proposition~4]{finster2026static}. Our factor-two adaptivity theorem therefore transfers this guarantee to the dynamic objective.

\section{Model and Benchmarks}
\label{sec:model}

We first describe the population and the outcome of a pooled test.  We then define histories, dynamic policies, and the two classes of static strategies used as benchmarks. Throughout, a negative test is the favorable outcome as it certifies that every member of the tested pool is healthy.

\paragraph{Population and feasible tests.}

\begin{definition}[Pooled-testing instance]
An instance is a tuple
\[
  J=\bigl([n],(u_i,q_i)_{i\in[n]},G\bigr).
\]
The agents are indexed by $[n]=\{1,\ldots,n\}$.  Agent $i$ has utility $u_i\ge0$ and an independent health indicator
\[
  X_i\sim\operatorname{Bernoulli}(q_i),
  \qquad X_i=1\text{ meaning healthy and }X_i=0\text{ meaning infected}.
\]
The integer $G\in\{1,\ldots,n\}$ is the maximum pool size.  Agents with $q_i=0$ can never be cleared and may be deleted, so we assume $q_i>0$.
\end{definition}

A \emph{feasible pool}, or simply a \emph{test}, is a nonempty set $t\subseteq[n]$ with $|t|\le G$.  We write
\[
  \mathcal T_G:=\{t\subseteq[n]:1\le |t|\le G\}
\]
for the finite set of feasible tests.

\begin{definition}[Test outcome]
For a health profile $x\in\{0,1\}^n$ and a pool $t\in\mathcal T_G$, define
\[
  r(t,x):=
  \begin{cases}
    -, & x_i=1\text{ for every }i\in t,\\
    +, & x_i=0\text{ for at least one }i\in t.
  \end{cases}
\]
Thus a test is negative exactly when all members of its pool are healthy.  A negative result \emph{clears} every agent in the pool whereas a positive result clears no one.
\end{definition}

\paragraph{Histories and realized welfare.}

\begin{definition}[Testing history]
A length-$b$ history is an ordered sequence
\[
  H_b=\bigl((t_1,r_1),\ldots,(t_b,r_b)\bigr)
  \in (\mathcal T_G\times\{-,+\})^b.
\]
It is \emph{feasible} if there is at least one profile $x\in\{0,1\}^n$ such that $r_j=r(t_j,x)$ for every $j\le b$.  Let $\mathcal H_b$ denote the set of feasible length-$b$ histories, and let $H_0=\varnothing$ and $\mathcal H_0=\{\varnothing\}$.
\end{definition}

The order in a history matters because later tests may depend on earlier outcomes.  The set of agents cleared by $H_b$ and the welfare realized along that history are given by
\[
  C(H_b):=\bigcup_{j:r_j=-}t_j,
  \qquad
  w(H_b):=\sum_{i\in C(H_b)}u_i.
\]
The union in $C(H_b)$ ensures that an agent's utility is counted only once, even if the agent appears in several negative pools.  

\paragraph{Dynamic policies.}

\begin{definition}[Dynamic policy]
Fix a test budget $B\ge0$.  A deterministic dynamic policy of horizon $B$ is a sequence of maps
\[
  \pi=(\tau_1,\ldots,\tau_B),
  \qquad
  \tau_{b+1}:\mathcal H_b\longrightarrow
  \mathcal T_G\cup\{\mathsf{stop}\},
  \quad b=0,\ldots,B-1.
\]
After observing history $H_b$, the policy either chooses the next pool $t_{b+1}=\tau_{b+1}(H_b)$ or stops.  If it chooses a pool, the outcome is observed and $(t_{b+1},r(t_{b+1},X))$ is appended to the history.  The process stops when the policy selects $\mathsf{stop}$ or when $B$ tests have been used.
\end{definition}

It is enough to define $\tau_{b+1}$ on histories reachable under the policy's earlier choices, but we define it on all of $\mathcal H_b$ for convenience. For a realized profile $x$, let $H_\pi(x)$ be the terminal history generated by this procedure. The policy's expected welfare is given by
\[
  V(\pi):=\E\bigl[w(H_\pi(X))\bigr]
  =\sum_{i=1}^n u_i\Prob_\pi\bigl(i\in C(H_\pi(X))\bigr).
\]
We focus throughout on deterministic dynamic policies, without loss of generality. Indeed, the internal random choices of any randomized policy can be sampled in advance, making it a distribution over deterministic policy trees. Its expected welfare is then simply an average of the welfare achieved by those trees and therefore cannot exceed the value of the best deterministic policy. Thus, all results stated for deterministic policies extend immediately to randomized ones.

\paragraph{Static strategies.}

\begin{definition}[Static overlapping strategy]
A static strategy with budget $B$ is a list
\[
  T=(t_1,\ldots,t_k),
  \qquad 0\le k\le B,
  \qquad t_j\in\mathcal T_G,
\]
chosen before any outcomes are observed where its pools may overlap. On profile $x$, it clears
\[
  C(T,x):=\bigcup_{j:r(t_j,x)=-}t_j,
\]
and its expected welfare is given by
\[
  V(T):=\E\left[\sum_{i\in C(T,X)}u_i\right].
\]
\end{definition}

The word \emph{allocation} will be used interchangeably with static strategy. Although the list notation records an order, that order does not affect $V(T)$ because every pool is selected in advance.

\begin{definition}[Static non-overlapping strategy]
A static strategy $T=(t_1,\ldots,t_k)$ is \emph{non-overlapping} if its pools are pairwise
disjoint: $t_j\cap t_\ell=\varnothing$ whenever $j\ne\ell$.
\end{definition}

Every static strategy can be viewed as a dynamic policy that ignores all observed outcomes and follows its preselected list.  Non-overlap is an additional restriction on a static strategy, not a restriction imposed by the testing technology.

\paragraph{Optimal benchmarks and adaptivity ratios.}

\begin{definition}[Optimal values]
For budget $B$ and instance $J$, define
\begin{align*}
  \OPTdyn(B,J)
    &:=\max_{\pi\text{ dynamic, horizon }B} V(\pi),\\
  \OPTstar(B,J)
    &:=\max_{T\text{ static, overlap allowed}} V(T),\\
  \OPTno(B,J)
    &:=\max_{T\text{ static, non-overlapping}} V(T).
\end{align*}
Because all three feasible classes are finite, these maxima are attained.  Their nesting gives
\[
  \OPTdyn(B,J)\ge \OPTstar(B,J)\ge \OPTno(B,J).
\]
\end{definition}

\begin{definition}[Gap ratios]
The \emph{adaptivity gap} and the dynamic-to-non-overlapping ratio are
\[
  \Agap(B,J):=\frac{\OPTdyn(B,J)}{\OPTstar(B,J)},
  \qquad
  \Ano(B,J):=\frac{\OPTdyn(B,J)}{\OPTno(B,J)},
\]
with either ratio defined to be one when its denominator is zero.
\end{definition}

\begin{remark}[Benchmark terminology]
Finster et al.\ study the purely static overlap ratio $\OPTstar/\OPTno$ and prove that it is at most two. By contrast, our adaptivity gap is $\OPTdyn/\OPTstar$: the static benchmark already permits overlapping pools, so this ratio isolates the additional value of conditioning future tests on observed outcomes.
\end{remark}

\paragraph{One-pool and fractional benchmarks.}

For a pool $t\in\mathcal T_G$, write
\[
  q_t:=\prod_{i\in t}q_i,
  \qquad
  u(t):=q_t\sum_{i\in t}u_i,
  \qquad
  u^*:=\max_{t\in\mathcal T_G}u(t).
\]
The quantity $q_t$ is the prior probability that $t$ tests negative, so $u(t)$ is the prior expected welfare from using a single test on $t$.  Also write
\[
  m_i:=u_iq_i,
\]
and order these numbers as $m_{(1)}\ge m_{(2)}\ge\cdots\ge m_{(n)}$, extending the sequence by $m_{(k)}=0$ for $k>n$.

For real $x\ge0$, define the fractional top sum
\begin{equation}
  T_x:=\sum_{k=1}^{\lfloor x\rfloor}m_{(k)}
       +(x-\lfloor x\rfloor)m_{(\lfloor x\rfloor+1)}.
  \label{eq:top-sum}
\end{equation}
Equivalently, $T_x$ is the optimum of
\[
  \max\left\{\sum_i m_ix_i:0\le x_i\le1,\ \sum_i x_i\le x\right\}.
\]
The map $x\mapsto T_x$ is nondecreasing and concave, with $T_0=0$.  Hence $T_x/x$ is nonincreasing on $x>0$, and for all $x,y>0$,
\begin{equation}
  T_x\le \max\{1,x/y\}T_y.
  \label{eq:top-sum-comparison}
\end{equation}

\paragraph{Clearance-spread parameter.}

The following statistic measures a population-wide bound on how widely one negative pool can spread clearance probability:
\begin{equation}
  \Lambda(J,G):=
  \max_{A\subseteq[n],\ |A|\le G}
  \sum_{i\in A}q_{A\setminus\{i\}}.
  \label{eq:lambda}
\end{equation}
For singleton $A = \{i\}$, we automatically have $|A| \leq G$ and the sum evaluates to 1, hence $\Lambda\ge1$. On the other hand, when $|A| = G$, each term in the sum is at most one, so $1\le\Lambda\le G$.

\subsection{A Strict Adaptivity Example}
\label{subsec:strict-example}

The following example shows that the adaptivity gap is not always one, i.e. there are examples where adaptivity provides a strict benefit.

\begin{example}
\label{ex:dynamic}
Consider three agents $\{A,B,C\}$, budget $B=2$, and pool cap $G=3$, with
\[
  (u_A,q_A)=(0.129,0.5562),\qquad
  (u_B,q_B)=(0.17483,1),\qquad
  (u_C,q_C)=(0.569,0.12).
\]
There are only seven nonempty feasible pools.  Direct enumeration of the resulting finite static and dynamic strategy spaces shows that an optimal static overlapping allocation is $(\{A,B\},\{B,C\})$. Its expected welfare is
\begin{align*}
  V_{\mathrm{stat}}
  &=u_Aq_A+u_Cq_C
    +u_B\bigl[1-(1-q_A)(1-q_C)\bigr]\\
  &=0.246580992.
\end{align*}
Here $A$ is cleared precisely when the first pool is negative, $C$ is cleared precisely when the second pool is negative, and the certainly healthy agent $B$ is cleared when at least one of those events occurs.

An optimal dynamic policy first tests $\{A,B\}$. Following a positive outcome it tests $\{B\}$; following a negative outcome it tests $\{C\}$.  Because $B$ is certainly healthy, the second test clears $B$ after a positive first outcome; after a negative first outcome, $A$ and $B$ are already cleared and the policy gives $C$ a singleton test.  The dynamic value is therefore
\begin{align*}
  V_{\mathrm{dyn}}
  &=u_Aq_A+u_B+u_Cq_Aq_C\\
  &=0.284557136.
\end{align*}
Consequently, for this instance, which we denote by $J_{\mathrm{ex}}$, and the two-test budget,
\[
  \Agap(2,J_{\mathrm{ex}})
  =\frac{0.284557136}{0.246580992}
  =1.154010831\ldots.
\]
For comparison, the optimal static non-overlapping allocation is $(\{A\},\{B\})$, with value
\[
u_Aq_A+u_Bq_B=0.2465798.
\]
Thus, in this example, permitting overlap improves the optimal static value only negligibly and the strict gap is driven almost entirely by adaptivity.
\end{example}

\section{Structural Theory}
\label{sec:structural-main}

This section develops the small set of structural facts used by our first two approximation results. Because these facts carry much of the intuition behind the results, we first explain the mechanism behind each one and then give its complete proof. Posterior normal forms, boundary regimes, hardness results, and greedy extensions are collected in \cref{app:technical-proofs}.

\subsection{Posterior Domination}

Positive pooled tests create dependence among agents, so marginal posterior probabilities can become complicated over the execution of a dynamic strategy. Our first key lemma allows us to compare joint clearance probabilities of tests during the execution of a dynamic strategy to original product priors. In what follows, we will repeatedly use the following event form of the product-measure correlation inequality
\citep{harris1960lower,fortuin1971correlation}.
\begin{harrisfkg}
Let $\mu=\bigotimes_{r\in K}\mu_r$ be a product probability measure on $\{0,1\}^K$, ordered coordinatewise.  If $\mathcal I$ and $\mathcal J$ are increasing events, then
\[
  \mu(\mathcal I\cap\mathcal J)\ge \mu(\mathcal I)\mu(\mathcal J).
\]
Equivalently, if $\mathcal I$ is increasing and $\mathcal D$ is decreasing, then
\[
  \mu(\mathcal I\cap\mathcal D)\le \mu(\mathcal I)\mu(\mathcal D),
\]
and, whenever $\mu(\mathcal D)>0$,
\[
  \mu(\mathcal I\mid\mathcal D)\le \mu(\mathcal I).
\]
\end{harrisfkg}

\begin{lemma}[Posterior domination]
\label{lem:main-posterior}
Let $H$ be a positive-probability history, let $N(H)$ be the agents already cleared by negative tests, and let $A\subseteq[n]\setminus N(H)$.  Then
\[
  \Prob(X_i=1\text{ for all }i\in A\mid H)
  \le \prod_{i\in A}q_i.
\]
In particular, no uncleared agent's posterior health probability exceeds its prior.
\end{lemma}

\paragraph{Intuition.}
Negative outcomes simply fix their members to be healthy therefore after removing those coordinates, the prior among remaining members is still a product distribution.  Each positive outcome on the other hand says that at least one member of its residual pool is infected, which is a decreasing event in the health coordinates. Moreover, the event that all agents in $A$ are healthy is increasing. As a result, the increasing--decreasing form of Harris--FKG bounds the conditional probability of the all-healthy event by its product-prior value.  

In addition, the conclusion is anchored at the original priors, not at the posterior associated with an earlier history. Posterior health probabilities need not decrease monotonically as test outcomes are observed: learning that one member of a previously positive pool is infected may “exonerate” the other members from previous positive tests they're included in and increase their posterior health probabilities (See Appendix \ref{sec:exoneration}). The result says that, despite such reversals, those probabilities can never rise above their original priors.

\begin{proof}
Write the negative pools in $H$ as $N_1,\ldots,N_r$ and the positive pools as $S_1,\ldots,S_k$. Conditioning on the negative outcomes pins the health indicator of every coordinate in
\[
  N:=\bigcup_{\ell=1}^r N_\ell=N(H)
\]
to one.  After these coordinates are removed, the remaining health indicators still follow their independent priors and we denote this product measure on the residual cube, $\{0,1\}^{[n]\setminus N}$, by $\mu$. Under $\mu$, all of the positive pool outcome information from $S_1,\ldots,S_k$ is captured by the event
\[
  E:=\bigcap_{\ell=1}^k
  \left\{\text{some }j\in S_\ell\setminus N\text{ has }X_j=0\right\}.
\]
Define the all-healthy event
\[
  \mathcal I_A:=\{X_i=1\text{ for every }i\in A\}.
\]
The event $\mathcal I_A$ is increasing on the residual cube, whereas $E$ is decreasing. Since $H$ is assumed to have positive probability, it follows that $\mu(E)>0$. Applying the increasing--decreasing Harris--FKG inequality with $\mathcal I=\mathcal I_A$ and $\mathcal D=E$ gives
\[
  \mu(\mathcal I_A\cap E)\le\mu(\mathcal I_A)\mu(E),
  \qquad\text{hence}\qquad
  \mu(\mathcal I_A\mid E)\le\mu(\mathcal I_A).
\]
Consequently,
\[
  \Prob(X_A=\mathbf1\mid H)
  =\mu(\mathcal I_A\mid E)
  \le \mu(\mathcal I_A)
  =\prod_{i\in A}q_i.
\]
Taking $A=\{i\}$ for any $i \notin N(H)$ gives the marginal conclusion.
\end{proof}

\subsection{Two Upper Bounds and Their Consequences}

Posterior domination yields two benchmarks that are repeatedly useful.  Recall the fractional top sum $T_x$, the one-pool optimum $u^*$, and the clearance-spread parameter $\Lambda(J,G)$ from \cref{sec:model}.

\begin{theorem}[Master upper bounds]
\label{thm:main-master-bounds}
Every dynamic policy $\pi$ with budget $B$ satisfies
\[
  V(\pi)\le T_{B\Lambda}
  \qquad\text{and}\qquad
  V(\pi)\le Bu^*.
\]
\end{theorem}

\paragraph{Intuition.}
For the first bound, assign to each agent the probability that it is first cleared, divided by its prior health probability. This theorem proceeds by showing that one test can add a total sum of clearance weights of at most $\Lambda$. The best possible weighted utility is therefore the fractional top sum $T_{B\Lambda}$.  For the second bound, decompose welfare by the first test that clears each agent.  Conditional on the history, the expected immediate welfare of that test is at most its prior value and hence at most $u^*$.

\begin{proof}
Fix a deterministic policy $\pi$. For bookkeeping, regard every unused
step on a branch that stops early as a formal empty test
$t_b=\varnothing$, declared to have a negative outcome. Such a test
clears no agent and contributes no welfare. We adopt the conventions
$q_\varnothing=1$ and $u(\varnothing)=0$.

Before step $b$, let $R_b$ denote the set of uncleared agents, and
define
\[
  A_b:=t_b\cap R_b.
\]
Thus $A_b$ is measurable with respect to the preceding history
$H_{b-1}$ and is empty on every padded step. For each agent $i$, define
the event
\[
  F_{i,b}
  :=\{i\in A_b\text{ and }t_b\text{ tests negative}\}.
\]
This is precisely the event that agent $i$ is first cleared at step
$b$. For each fixed $i$, the events
$F_{i,1},\ldots,F_{i,B}$ are disjoint, and their union is the event
that $i$ is cleared. We may therefore define the normalized clearance
probability
\[
  \rho_i
  :=\frac{\Prob(i\text{ is cleared})}{q_i}
  =\frac{1}{q_i}\sum_{b=1}^B\Prob(F_{i,b}).
\]
An agent can be cleared only when healthy, so
\[
  \Prob(i\text{ is cleared})
  \le \Prob(X_i=1)
  =q_i.
\]
It follows that $0\le\rho_i\le1$.

We first prove the bound involving $T_{B\Lambda}$. Fix a step $b$ and a realization of the preceding history $H_{b-1}$ having positive probability. The sets $t_b$ and $A_b$ are then fixed. Moreover, every member of $t_b\setminus A_b$ has already been cleared and is therefore known to be healthy. Consequently, $t_b$ tests negative exactly when every agent in $A_b$ is healthy. Since $A_b$ contains only uncleared agents, posterior domination gives
\[
  \Prob(t_b\text{ tests negative}\mid H_{b-1})
  =\Prob(X_{A_b}=\mathbf1\mid H_{b-1})
  \le \prod_{j\in A_b}q_j
  =q_{A_b}.
\]
This also holds on padded steps, since then both events in the equality are certain and $q_{A_b}=q_\varnothing=1$.

Because $A_b$ is measurable with respect to $H_{b-1}$, the law of total expectation gives
\begin{align*}
  \Prob(F_{i,b})
  &=
  \E\left[
    \Prob(F_{i,b}\mid H_{b-1})
  \right]\\
  &=
  \E\left[
    \mathbf1\{i\in A_b\}
    \Prob(t_b\text{ tests negative}\mid H_{b-1})
  \right].
\end{align*}
Summing this identity over $i$ and applying the preceding posterior bound yields
\begin{align*}
  \sum_i\frac{\Prob(F_{i,b})}{q_i}
  &=
  \E\left[
    \Prob(t_b\text{ tests negative}\mid H_{b-1})
    \sum_{i\in A_b}\frac{1}{q_i}
  \right]\\
  &\le
  \E\left[
    q_{A_b}\sum_{i\in A_b}\frac{1}{q_i}
  \right]\\
  &=
  \E\left[
    \sum_{i\in A_b}q_{A_b\setminus\{i\}}
  \right]\\
  &\le \Lambda.
\end{align*}
The expectations are over the random history $H_{b-1}$ induced by the prior and the policy. The final inequality holds pointwise: when $A_b$ is nonempty, it is a feasible pool and the definition of $\Lambda$ applies; when $A_b=\varnothing$, the sum vanishes. Thus the bound also covers padded steps.

Summing over $b=1,\ldots,B$ gives
\[
  \sum_i\rho_i
  =\sum_{b=1}^B
    \sum_i\frac{\Prob(F_{i,b})}{q_i}
  \le B\Lambda.
\]
Consequently,
\[
  V(\pi)
  =\sum_i u_i\Prob(i\text{ is cleared})
  =\sum_i u_iq_i\rho_i
  =\sum_i m_i\rho_i
  \le T_{B\Lambda}.
\]
The final inequality follows from the fractional-knapsack characterization of $T_{B\Lambda}$, since $0\le\rho_i\le1$ for every $i$ and
$\sum_i\rho_i\le B\Lambda$.

For the second bound, decompose welfare according to the step at which each agent is first cleared:
\[
  V(\pi)
  =\sum_{b=1}^B
    \E\left[
      \mathbf1\{t_b\text{ tests negative}\}
      \sum_{i\in A_b}u_i
    \right].
\]
Fix a realization of $H_{b-1}$ having positive probability. As observed above, $t_b$ tests negative exactly when every agent in $A_b$ is healthy. Posterior domination therefore gives
\begin{align*}
  &\E\left[
    \left.
    \mathbf1\{t_b\text{ tests negative}\}
    \sum_{i\in A_b}u_i
    \,\right|\,H_{b-1}
  \right]\\
  &\qquad=
  \Prob(X_{A_b}=\mathbf1\mid H_{b-1})
  \sum_{i\in A_b}u_i\\
  &\qquad\le
  q_{A_b}\sum_{i\in A_b}u_i\\
  &\qquad=
  u(A_b)
  \le u^*.
\end{align*}
When $A_b=\varnothing$, the conditional contribution is zero, so the same bound remains valid on padded steps. Taking expectations, summing over $b=1,\ldots,B$, and using the preceding welfare decomposition gives
\[
  V(\pi)\le Bu^*.
\]
\end{proof}

\begin{theorem}[Boundary-sensitive adaptivity bounds]
\label{thm:main-boundaries}
The following refinements hold.
\begin{enumerate}[label=(\roman*),leftmargin=*]
\item If $\Lambda(J,G)\le1$, then
\[
  \OPTdyn(B,J)=\OPTstar(B,J)=\OPTno(B,J)
  =\sum_{b=1}^{\min\{B,n\}}m_{(b)}.
\]
The common optimum tests individually the agents with the largest $m_i=u_iq_i$.  In particular, this conclusion holds if every $q_i\le1/2$.
\item If $B=1$ or $B\ge n$, then $\Agap(B,J)=1$.
\item Suppose there are $B$ pairwise disjoint feasible pools $t_1,\ldots,t_B$ with $u(t_b)\ge(1-\gamma)u^*$ for every $b$, where $0\le\gamma<1$.  Then
\[
  \Agap(B,J)\le\Ano(B,J)\le\frac1{1-\gamma}.
\]
If the best one-pool value can be replicated exactly and disjointly $B$ times, all three optimal values equal $Bu^*$.
\item Suppose $q_i\ge\theta\in(0,1]$ for every agent and $n\ge BG$.  Then
\begin{align}
  \Agap(B,J)
  &\le \min\!\left\{2,\Ano(B,J)\right\} \notag\\
  &\le \min\!\left\{
      2,
      \min_{1\le s\le G}
      \frac{\max\{1,\Lambda(J,G)/s\}}{\theta^{s-1}}
    \right\} \notag\\
  &\le \min\{2,\phi_G(\theta)\},
  \qquad
  \phi_G(\theta):=\min_{1\le s\le G}\frac{G}{s\theta^{s-1}}.
  \label{eq:main-high-q}
\end{align}
\end{enumerate}
Proofs are given in \cref{sec:boundaries}.
\end{theorem}

The high-health expression in \cref{eq:main-high-q} approaches one as $\theta\to1$.  It is the dynamic analogue of the high-health static overlap bound in \citet{finster2026static}; the extra condition $n\ge BG$ ensures that the $B$ comparison pools used here can be filled disjointly. \Cref{tab:main-high-q} reports the prior-only bound $\min\{2,\phi_G(\theta)\}$. The $\Lambda$-dependent middle expression in \cref{eq:main-high-q} can be smaller still.

\begin{table}[t]
  \centering
  \footnotesize
  \setlength{\tabcolsep}{4.2pt}
  \caption{Boundary-sensitive upper bounds on the adaptivity gap when $q_i\ge\theta$ and
  $n\ge BG$.  Each entry is $\min\{2,\phi_G(\theta)\}$.}
  \label{tab:main-high-q}
  \begin{tabular}{c@{\quad}rrrrrrr}
    \toprule
    $G$ & $\theta=.50$ & $.60$ & $.70$ & $.80$ & $.85$ & $.90$ & $.95$ \\
    \midrule
    2  & 2.00 & 1.67 & 1.43 & 1.25 & 1.18 & 1.11 & 1.05 \\
    3  & 2.00 & 2.00 & 2.00 & 1.56 & 1.38 & 1.24 & 1.11 \\
    4  & 2.00 & 2.00 & 2.00 & 1.95 & 1.63 & 1.37 & 1.17 \\
    5  & 2.00 & 2.00 & 2.00 & 2.00 & 1.92 & 1.52 & 1.23 \\
    8  & 2.00 & 2.00 & 2.00 & 2.00 & 2.00 & 2.00 & 1.43 \\
    10 & 2.00 & 2.00 & 2.00 & 2.00 & 2.00 & 2.00 & 1.59 \\
    \bottomrule
  \end{tabular}
\end{table}

\paragraph{Computational implications.}
Exact optimization is already hard in restricted cases.  For every fixed $G\ge3$, deciding whether the dynamic optimum exceeds a threshold is NP-hard.  Evaluating a supplied policy is \#P-hard (with a PP-hard threshold problem), and computing an optimal first test is W[1]-hard when parameterized by $G$.  These implications largely inherit the static complexity landscape of \citet{finster2026static}: policy evaluation and first-test optimization contain their overlapping evaluation and single-test problems as special cases, while our dynamic NP-hardness proof adapts their \textsc{3-Partition} construction and uses \cref{thm:main-master-bounds} to control adaptive policies.  Statements and reductions appear in \cref{sec:complexity}.  These hardness results motivate the greedy policies studied next and the exact small-instance experiments in \cref{sec:experiments}.

\subsection{Where Adaptivity Can Help}
\label{sec:where-adaptivity-helps}

Once an agent has been cleared by a negative test, the policy knows that the agent is healthy. Including that agent in a later pool therefore cannot affect the pool’s outcome or generate any additional welfare. The following observation formalizes why such agents may be removed from all subsequent pools without loss.

\begin{definition}[Normalized policy]
\label{def:normalized-policy}
A dynamic policy $\pi$ is \emph{normalized} if, at every reachable history $H$, each pool $t$ selected by $\pi$ satisfies
\[
t\cap C(H)=\varnothing.
\]
Thus, once an agent is cleared, that agent never appears in a later test.
\end{definition}

\begin{lemma}[Normalization without loss]
\label{lem:normalize-policy}
For every deterministic dynamic policy $\pi$ with budget $B$, there exists a normalized deterministic policy $\pi^{\mathrm{norm}}$ that uses at most $B$ tests and has the same realized welfare as $\pi$ on every infection profile. Consequently,
\[
V(\pi^{\mathrm{norm}})=V(\pi),
\]
and the optimization defining $\OPTdyn(B,J)$ may be restricted to normalized policies.
\end{lemma}

\begin{proof}
We transform the decision tree of $\pi$ recursively. Consider a reachable node corresponding to a history $H$, let $C=C(H)$ be the set of agents already cleared, and suppose that $\pi$ selects the pool $t$. Replace $t$ by
\[
t':=t\setminus C.
\]
Every agent in $C$ is known to be healthy conditional on $H$. Therefore, for every infection profile consistent with $H$, the pools $t$ and $t'$ have the same outcome. Moreover, if that outcome is negative, then
\[
C\cup t=C\cup t',
\]
so the two tests produce the same updated cleared set; if the outcome is positive, neither test clears a new agent.

If $t'\neq\varnothing$, replace $t$ by $t'$ and recursively apply the same transformation to its two continuation subtrees. If $t'=\varnothing$, then every member of $t$ has already been cleared. The outcome of $t$ is therefore deterministically negative and the test clears no new agent, so we contract the node to its negative continuation.

Applying this transformation throughout the finite decision tree yields a normalized policy that uses no more tests than $\pi$. On every infection profile, the transformed policy encounters the corresponding sequence of outcomes and clears exactly the same agents as $\pi$. Hence the two policies have the same realized welfare.
\end{proof}

Normalization removes only agents already known to be healthy. It does not prohibit a policy from re-pooling an unresolved agent following a positive test. In what follows, call a policy \emph{path-disjoint} if the pools on every root-to-leaf path are pairwise disjoint.

\begin{proposition}[No value from path-disjoint adaptation]
\label{prop:main-path-disjoint}
The optimal path-disjoint dynamic policy has value exactly $\OPTno(B,J)$. Consequently, any policy that beats the static non-overlapping optimum must, on some path, re-pool an agent from an earlier positive test.
\end{proposition}

\paragraph{Intuition.}
Once the root pool is fixed, both continuation policies are confined to the remaining agents, whose distribution is independent of the root outcome. One can keep the better continuation and obtain a static non-overlapping allocation of the same or greater value. Induction on the budget formalizes this replacement.

\begin{proof}
We induct on $B$. The claim is immediate for $B=0$. For $B\geq 1$, let $t$ be the root pool of a path-disjoint policy $\pi$, and let $\pi^-$ and $\pi^+$ be its continuations after a negative and positive outcome, respectively. Path-disjointness confines both continuations to $[n]\setminus t$. The health indicators of those remaining agents are independent of the root outcome, so the two continuation values may be evaluated under the same original product prior on $[n]\setminus t$. Hence
\begin{align*}
V(\pi)
&=
u(t)+q_tV(\pi^-)+(1-q_t)V(\pi^+)\\
&\leq
u(t)+\max\{V(\pi^-),V(\pi^+)\}\\
&\leq
u(t)+\OPTno\bigl(B-1,J|_{[n]\setminus t}\bigr),
\end{align*}
where $J|_{[n]\setminus t}$ denotes the induced instance on the remaining agents, and the final inequality is the induction hypothesis.

The last expression is attained by the static non-overlapping strategy consisting of $t$ together with an optimal $(B-1)$-test non-overlapping strategy on the remaining population. Consequently, no path-disjoint dynamic policy outperforms $\OPTno(B,J)$. The reverse inequality holds because every static non-overlapping strategy is itself path-disjoint.

For the consequence, suppose a policy never re-pools an agent after an earlier positive test. By \cref{lem:normalize-policy}, replace it by a normalized policy with the same value. In the normalized policy, an agent cannot reappear after belonging to a negative test, because that agent has already been cleared. By assumption, an agent also cannot reappear after belonging to a positive test. Hence no agent appears twice on a root-to-leaf path, so the normalized policy is path-disjoint. It follows that its value is at most $\OPTno(B,J)$. Therefore, any policy whose value exceeds $\OPTno(B,J)$ must re-pool an agent from an earlier positive test on some path.
\end{proof}

This identifies a source of adaptive value: positive tests create ambiguity, and re-pooling can use later tests to resolve it. The mechanism is present in the strict example of \cref{ex:dynamic}.

\section{Greedy Policies and Approximation}
\label{sec:greedy-main}

\subsection{Exact Conditional Greedy}

At history $H$, let $C(H)$ be the cleared set and $U(t,H)=t\setminus C(H)$ the uncleared members of a candidate pool.  The true expected immediate gain is
\begin{equation}
  g(t\mid H)
  :=\Prob(X_{U(t,H)}=\mathbf1\mid H)
    \sum_{i\in U(t,H)}u_i.
  \label{eq:main-exact-gain}
\end{equation}

\paragraph{Exact-Joint Greedy.}
$\Greedy$ selects a feasible pool maximizing \cref{eq:main-exact-gain}, observes the outcome, and repeats.

\paragraph{Marginal-Product Greedy.}
$\MPGreedy$ replaces the joint probability in \cref{eq:main-exact-gain} by
\[
  \prod_{i\in U(t,H)}\Prob(X_i=1\mid H).
\]
The two rules agree under an independent posterior, including at the root, but may diverge after a positive outcome.  The product of posterior marginals may lie above or below the joint probability, so the theorem below applies only to $\Greedy$.

The greedy analysis rests on a local stability property of each selected pool. This property parallels the split-stability condition for static allocations in \citet{finster2026static}, but its dynamic formulation is different: it holds conditional on an arbitrary history and combines exact one-step optimality with posterior domination.

\begin{lemma}[Conditional split stability]
\label{lem:main-split-stability}
Fix a positive-probability history $H$, and let $t$ be a pool selected by $\Greedy$ with $g(t\mid H)>0$.  Write $U:=U(t,H)$ and $\Sigma:=\sum_{h\in U}u_h$.  Then
\begin{equation}
  u_i\ge(1-q_i)\Sigma\quad\text{for every }i\in U,
  \qquad\text{and consequently}\qquad
  \sum_{i\in U}(1-q_i)\le1.
  \label{eq:main-split-stability}
\end{equation}
\end{lemma}

\paragraph{Intuition.}
If deleting agent $i$ made the selected pool more valuable, exact greedy would have selected the smaller pool instead. Thus the utility supplied by $i$ must compensate for the extra failure risk created by requiring $i$ to be healthy. Harris--FKG bounds that conditional health probability by the original prior $q_i$ which gives the first inequality in \ref{eq:main-split-stability}. Summing the resulting compensation inequalities shows that the selected pool contains at most one unit of total prior infection risk, the second inequality in \ref{eq:main-split-stability}.

\begin{proof}
Previously cleared members may be deleted from $t$ without changing its outcome or gain, so take $t=U$. The positive gain, $g(t\mid H)>0$, implies $\Sigma>0$.  If $|U|=1$, then $u_i=\Sigma\ge(1-q_i)\Sigma$ and the conclusion is immediate, so suppose $|U|\ge2$.

For every $i\in U$, the pool $U\setminus\{i\}$ is feasible, and exact maximization in the greedy step gives
\[
  g(U\mid H)\ge g(U\setminus\{i\}\mid H).
\]
Define
\[
  c_i:=
  \frac{\Prob(X_U=\mathbf1\mid H)}
       {\Prob(X_{U\setminus\{i\}}=\mathbf1\mid H)}
  =\Prob(X_i=1\mid H,\ X_{U\setminus\{i\}}=\mathbf1).
\]
Given we assume $g(t\mid H)>0$, the numerator is positive, so the denominator is positive as well. We next verify $c_i\le q_i$. Let $N=N(H)$, write the positive pools recorded in $H$ as $S_1,\ldots,S_m$, and let $\mu_i$ be the product measure obtained from the prior after pinning every coordinate in $N\cup(U\setminus\{i\})$ healthy (as in what we condition on for $c_i$). Under $\mu_i$, the information from positive tests on uncleared individuals is governed by the event
\[
  E_i:=\bigcap_{\ell=1}^m
  \left\{\text{some }r\in
  S_\ell\setminus\bigl(N\cup(U\setminus\{i\})\bigr)
  \text{ has }X_r=0\right\}.
\]
This event is decreasing and the event $\mathcal I_i:=\{X_i=1\}$ is increasing (both in the residual hypercube obtained by pinning every coordinate in $N\cup(U\setminus\{i\})$ as healthy), and the positivity of the denominator in the expression for $c_i$ gives $\mu_i(E_i)>0$.  The increasing--decreasing Harris--FKG inequality therefore gives
\[
  \mu_i(\mathcal I_i\cap E_i)
  \le\mu_i(\mathcal I_i)\mu_i(E_i),
  \qquad\text{so}\qquad
  c_i=\mu_i(\mathcal I_i\mid E_i)
  \le\mu_i(\mathcal I_i)=q_i.
\]

Using the fractional form of $c_i$ and comparing the selected pool with the pool obtained by dropping $i$ yields
\[
  c_i\Sigma\ge \Sigma-u_i,
  \qquad\text{hence}\qquad
  u_i\ge(1-c_i)\Sigma\ge(1-q_i)\Sigma.
\]
Summing over $i\in U$ and dividing by $\Sigma>0$ proves
\cref{eq:main-split-stability}.
\end{proof}

\begin{theorem}[Exact-Joint Greedy]
\label{thm:main-greedy}
For every population, budget, and pool cap,
\[
  V(\Greedy)\ge\frac{1}{e+1}\OPTstar(B,J).
\]
\end{theorem}

\paragraph{Intuition.}
Recall $\Greedy$ sequentially implements tests $g_1,\dots,g_B$. We let $R_1 = [n]$ and $R_{j+1} = R_j \setminus t_j$ such that $R_j$ represents the individuals not yet tested under $\Greedy$ at the beginning of the $j$-th step of its execution. The proof follows a sequence of static benchmarks, $F_1,\dots,F_{B+1}$ where $F_j = \OPTstar(B-j+1,J|_{R_j})$ is the optimal static welfare achievable with $B-j+1$ tests on the population restricted to $R_j$ (denoted by $J|_{R_j}$). The main step of the proof bounds the residual change in this benchmark with respect to utility obtained by $\Greedy$ at step $j$: $F_j - F_{j+1} \leq (e+1)g_j$. With this in hand, we can telescope the sum to show that $\OPTstar(B,J) = F_1 -F_{B+1} \leq (1+e) \sum_{j=1}^B g_j$. The theorem follows by taking expectations over the random realizations of $g_1,\dots,g_B$. 

\begin{proof}
Fix a history $H$ immediately before greedy step $j$. Let $t_j$ be the selected pool, write
\[
  U_j:=U(t_j,H),
  \qquad
  g_j:=g(t_j\mid H),
\]
let $R_j$ be the agents who have never appeared in an earlier test, and let
$D_j:=t_j\cap R_j$ be the fresh members of the selected pool.

\smallskip
\noindent\emph{Step 1: bounding total risk from $D_j$ (when $g_j>0$).}

\noindent
We begin by invoking conditional split stability to show a bound on the total infection risk arising from individuals in $D_j$. For $g_j > 0$, we can apply \cref{lem:main-split-stability} and the fact that $D_j \subseteq U_j$ to obtain: 
\begin{equation}
  \sum_{i\in D_j}(1-q_i)
  \le\sum_{i\in U_j}(1-q_i)\le1.
  \label{eq:main-fresh-risk}
\end{equation}

\smallskip
\noindent\emph{Step 2: bounding maximal extractable utility from $D_j$.}

\noindent
In this step we use the bound from Step 1 to show a bound on the maximal utility extractable from fresh agents, $\sum_{i \in D_j} q_iu_i \leq e g_j$. To begin, every feasible pool $S\subseteq R_j$ consists entirely of never-tested agents. Its health states are therefore independent of $H$, so its conditional gain equals its prior value
\[
  u(S):=q_S\sum_{i\in S}u_i.
\]
Exact greedy optimality consequently implies
\begin{equation}
  u(S)\le g_j
  \qquad(\text{for all }S\subseteq R_j,\ |S|\le G).
  \label{eq:main-fresh-comparison}
\end{equation}

If $g_j=0$, applying this comparison to every fresh singleton implies $q_i u_i = 0$ for all $i \in D_j$, hence $\sum_{i\in D_j}q_i u_i=0 = eg_j$ as desired. Consequently, we proceed by assuming that $g_j>0$. Form an auxiliary random subset $A\subseteq D_j$ made by including each $i$ independently with probability $q_i$ (This subset is only a charging device, not a test performed by the policy). As before, let $q_A:=\prod_{i\in A}q_i$ and $q_\varnothing:=1$ so that via \cref{eq:main-fresh-comparison}, each realization of $A$ gives
\[
  \sum_{i\in A}u_i\le \frac{g_j}{q_A}.
\]
For $A=\varnothing$ the left side is zero, so the inequality remains valid.  Taking expectations over this auxiliary sampling
\[
  \sum_{i\in D_j}q_i u_i
  =\sum_{i\in D_j}\Pr(i\in A) u_i
  = \E\left[\sum_{i\in A}u_i\right]
  \leq g_j\E\left[q_A^{-1}\right].
\]
To compute the remaining expectation, let $B_i:=\1\{i\in A\}$.  The variables $B_i$ are
independent with $B_i\sim\operatorname{Bernoulli}(q_i)$, and
\[
  q_A^{-1}=\prod_{i\in D_j}q_i^{-B_i}.
\]
Using independence coordinate by coordinate,
\begin{align*}
  \E[q_A^{-1}]
  &=\prod_{i\in D_j}\E[q_i^{-B_i}]\\
  &=\prod_{i\in D_j}
    \left[(1-q_i)\cdot1+q_i\cdot q_i^{-1}\right]\\
  &=\prod_{i\in D_j}(2-q_i).
\end{align*}
It follows that
\begin{align}
  \sum_{i\in D_j}q_i u_i \le g_j\prod_{i\in D_j}(2-q_i)
    \le g_j\exp\left(\sum_{i\in D_j}(1-q_i)\right)
    \le e g_j.                                   \label{eq:main-fresh-charge}
\end{align}
The penultimate inequality uses $2-q_i=1+(1-q_i)\le e^{1-q_i}$ and the final inequality is
\cref{eq:main-fresh-risk}.

\smallskip
\noindent\emph{Step 3: bounding $F_{j} - F_{j+1}$.}

\noindent
Along the realized greedy path, let
\[
  F_j:=\OPTstar(B-j+1,J|_{R_j}),
  \qquad
  R_{j+1}:=R_j\setminus t_j=R_j\setminus D_j,
\]
where $J|_{R_j}$ is the induced instance on the never-tested agents.  Fix an optimal static allocation attaining $F_j$.  Delete every agent in $D_j$ from all of its pools and discard any empty pools. Removing companions cannot hurt the remaining agents. The total welfare lost from the deleted agents is at most their prior singleton welfare ($\sum_{i \in D_j} q_iu_i$), and therefore at most $e g_j$ by
\cref{eq:main-fresh-charge}.

The resulting allocation may still use $B-j+1$ pools, whereas the benchmark $F_{j+1}$ allows only $B-j$.  If necessary, delete one remaining pool $s$.  Its marginal contribution to an overlapping allocation is at most its stand-alone prior value $u(s)$: deleting the pool can only lose agents whose clearance depended on that pool. Since $s\subseteq R_{j+1}\subseteq R_j$, \cref{eq:main-fresh-comparison} bounds this additional loss by $g_j$. We have constructed a feasible allocation for $F_{j+1}$ and proved, pathwise,
\begin{equation}
  F_j-F_{j+1}\le(e+1)g_j.
  \label{eq:main-greedy-telescope}
\end{equation}
Finally, $R_1=[n]$ and $F_{B+1}=0$, summing \cref{eq:main-greedy-telescope} over the greedy steps gives
\[
  \OPTstar(B,J)=F_1\le(e+1)\sum_{j=1}^B g_j.
\]

Finally, take expectations over the random history generated by Greedy. Let $Y_j$ denote the welfare actually gained at step $j$. By definition, $g_j=\E[Y_j\mid H_{j-1}]$. Hence the law of iterated expectations gives
\[
  \E\left[\sum_j g_j\right]
  =\sum_j\E[Y_j]
  =V(\Greedy),
\]
which proves the theorem.

\end{proof}

Extending the argument to approximate joint-probability oracles requires more than just approximately maximizing immediate gains to select a pool: the oracle must also preserve the requisite subset stability within that pool. The resulting conditional guarantee is given in \cref{sec:approx-oracle}.

\subsection{Exact Dynamic Greedy Versus Static Greedy}
\label{subsec:greedy-comparison}

For comparison, define $\StaticGreedy$ to be the exact-oracle algorithm of
\citet{finster2026static}: starting from the full population, it repeatedly chooses a feasible pool with maximum prior value $u(t)=q_t\sum_{i\in t}u_i$, assigns that pool as the next static test, and removes all of its members before the next round. Its output is therefore static and non-overlapping.  Finster et al. prove
\begin{equation}
  V(\StaticGreedy)\ge\frac1{e+1}\OPTstar(B,J).
  \label{eq:static-greedy-finster}
\end{equation}
Thus $\StaticGreedy$ and $\Greedy$ have the same certified $1/(e+1)$ fraction of the static overlapping benchmark. Combining \cref{eq:static-greedy-finster,thm:main-greedy} with the factor-two theorem later in the paper gives each algorithm a $1/[2(e+1)]$ fraction of optimal dynamic welfare, equivalently a $2(e+1)$-approximation; see \cref{cor:greedy-dynamic}.

Although the two algorithms satisfy the same benchmark guarantee, their pointwise relationship remains open. $\StaticGreedy$ commits to a collection of disjoint pools in advance, whereas $\Greedy$ recomputes joint posterior gains after every outcome and may re-pool agents following positive tests. This adaptive flexibility provides additional options, but $\Greedy$ remains a myopic rule, so the existing analysis does not imply pointwise dominance.

The exact experiments in \cref{sec:experiments} nevertheless reveal a consistent pattern: across all 600 sampled instances, $\Greedy$ was weakly better than $\StaticGreedy$, with the head-to-head frequencies reported in \cref{tab:greedy-head-to-head}. This evidence motivates the conjecture that $\Greedy$ may dominate $\StaticGreedy$ pointwise, but finite computation cannot establish such a result. We leave proving this conjecture or constructing a counterexample as an important direction for future work.

\begin{table}[ht]
  \centering
  \small
  \caption{Preliminary exact comparison of the two greedy algorithms.  Entries are percentages of sampled instances; equality is determined using the numerical tolerance of the exact solver.}
  \label{tab:greedy-head-to-head}
  \begin{tabular}{lrrrr}
    \toprule
    $(n,B,G)$ & Instances & $\Greedy>\StaticGreedy$ & Tie & $\StaticGreedy>\Greedy$ \\
    \midrule
    $(4,2,4)$ & 300 & 38.3\% & 61.7\% & 0.0\% \\
    $(5,3,5)$ & 200 & 52.5\% & 47.5\% & 0.0\% \\
    $(6,3,3)$ & 100 & 71.0\% & 29.0\% & 0.0\% \\
    \bottomrule
  \end{tabular}
\end{table}

\section{A Uniform Factor-two Adaptivity Gap}
\label{sec:factor-two}

In this section, we prove our main result: for every pooled-testing instance and every test budget, the optimal dynamic welfare is at most twice the optimal static overlapping welfare. Equivalently,
\[
  \OPTdyn(B,J)\le 2\OPTstar(B,J).
\]
Fix a deterministic dynamic policy $\pi$. It is enough to construct a distribution over feasible static allocations whose expected welfare is at least $V(\pi)/2$, since the best deterministic static allocation must be at least as valuable as the average allocation in that distribution.

A tempting construction is to sample a virtual infection profile, run $\pi$ on that profile, and use the pools encountered along the resulting path as a static allocation. This produces a feasible random static allocation, which can then be evaluated on an independent infection profile. The difficulty is that the pools on a dynamic path are selected in response to the outcomes generated by the virtual profile. Their selection is therefore correlated with the very information that makes them valuable dynamically. Once the recorded pools are evaluated on a fresh profile, this correlation disappears: an infected companion can prevent an entire pool from clearing, and the unmodified path allocation need not retain any constant fraction of the dynamic policy's value.

The solution is to \emph{half-thin} the sampled path allocation: independently for every pool on the path, retain each agent in that pool with probability $1/2$. Suppose we focus on whether a given agent (a focal agent) is cleared in this process. Thinning may discard this focal agent whose clearance we wish to preserve, but it may also discard infected companions who would otherwise prevent the focal agent from being cleared. The proof shows that these competing effects balance in aggregate. For each focal agent, a Boolean-tree charging argument compares the profiles on which the dynamic policy clears that agent with the probability that one of the corresponding thinned pools clears it on a complementary profile. The resulting loss is at most a factor of two.

We first establish this comparison on a uniform Boolean subcube, where free infection coordinates can be paired with their complements. The core of this argument reduces the portion of the policy relevant to one focal agent to a query--offer model, defined formally below: tests that exclude the focal agent act as \emph{queries} that determine the path, whereas tests that include it act as \emph{offers} that may clear it. Our key combinatorial lemma analyzes this reduced tree. Finally, a two-profile coupling conditions arbitrary heterogeneous product priors into uniform Boolean subcubes. Averaging the resulting per-agent inequalities shows that the half-thinned path allocation has expected welfare at least $V(\pi)/2$, completing the comparison with $\OPTstar(B,J)$.

\paragraph{Proof strategy.}
The proof compares the dynamic policy and the random static allocation one agent at a time by fixing a focal agent $i$. We use one infection profile to select a path through the dynamic policy, half-thin the pools on that path, and evaluate the resulting static allocation on an independent infection profile. The central step is to prove that the probability this static allocation clears $i$ is at least half the probability that the original dynamic policy clears $i$. Once this inequality is established for every agent, we multiply by the utilities and sum. Every realization of the half-thinned path is a feasible static allocation, so its expected welfare is bounded by $\OPTstar(B,J)$, yielding the factor of two.

Most of the work lies in proving the per-agent comparison. We begin on a uniform Boolean subcube, where infection profiles can be paired by flipping their free coordinates. From the perspective of a focal agent, $i$, a test that excludes $i$ only helps determine which part of the policy is reached, while a test containing $i$ provides an opportunity to clear it. The query--offer tree introduced below formalizes this distinction and supports the required combinatorial argument. A two-profile coupling then reduces arbitrary heterogeneous product priors to these uniform-subcube comparisons. The following subsections develop each of these steps formally.

\paragraph{Notation and half-thinning.}
Throughout this section, it is more convenient to use infection
indicators
\[
  Z_i:=1-X_i,
  \qquad Z_i=1\text{ meaning infected}.
\]
For a deterministic policy $\pi$, write $P_\pi(z)$ for the list of at most $B$ pools encountered when the infection profile is $z$. Given any list of pools $P$, its \emph{half-thinning} independently retains each agent--pool incidence with probability $1/2$, using coins independent of all infection profiles. Thus, if an agent appears in several pools, its retention is decided independently in each occurrence. Empty resulting pools are discarded. Because thinning can only decrease pool sizes, every half-thinned list is a feasible static overlapping allocation using at most $B$ tests.

\subsection{Reduced Normalized Tree}

This section isolates the main combinatorial object at the heart of the proof. Let $E \subseteq [n]$ be a finite set of free Boolean coordinates, order $\{0,1\}^E$ coordinatewise, and for $z \in \{0,1\}^E$, denote the support of an infection profile by  $\operatorname{supp}(z):=\{j\in E:z_j=1\}$. In addition we let $\bar{z} \in \{0,1\}^E$ be the \emph{companion profile} to $z$ such that $\bar{z}_i = 1 - z_i$.  

The object below is a bookkeeping device for comparing the two clearance events at the center of the proof. Fix a focal agent \emph{assumed to be healthy} and imagine restricting the infection profiles to a Boolean subcube, so that the coordinates outside $E$ are already fixed. After the effects of those fixed coordinates are removed, a profile $z\in\{0,1\}^E$ determines the path followed through the dynamic policy (i.e. tests made). At the same time, every opportunity to clear the healthy focal agent is equipped with an independent thinning trial that records the probability that the corresponding half-thinned pool clears the focal agent when evaluated on the companion profile $\bar{z}$. Roughly, tests that exclude the focal agent become \emph{queries}, because they only determine which path is followed, while tests containing the focal agent become \emph{offers}, because they provide opportunities to clear it. Thus, the same tree tracks both dynamic clearance under the path-selecting profile $(z)$ and randomized static clearance under its companion profile $(\bar{z})$.

\begin{definition}[Reduced normalized tree]
Fix a finite set $E$ of free nonfocal coordinates, together with an implicit distinguished healthy focal agent not represented in $E$. A \emph{reduced normalized tree} on $E$ is a finite rooted tree, denoted by $T$,  whose leaves are unlabeled and whose internal nodes are of the following two
types:

\begin{enumerate}[label=(\roman*),leftmargin=*]

\item A \emph{query node} (test without focal agent) is labelled by a set $A\subseteq E$ and has two children, called its zero child and its positive child. For a profile $z\in\{0,1\}^E$, the execution takes the zero child when
\[
  z_A=\mathbf0,
\]
meaning that $z_j=0$ for every $j\in A$, and takes the positive child otherwise. The tree is \emph{normalized} in the sense that every query or offer label appearing in the subtree below the zero child is disjoint from $A$.

\item An \emph{offer node} (test with focal agent) is labelled by a set $S\subseteq E$ and has one continuation child. When the execution reaches this node, an independent thinning trial is performed. Conditional on the profile $z$, this trial succeeds with probability
\begin{equation}
  p_S(z):=
  \left(\frac{1}{2}\right)^{1+|S \cap\operatorname{supp}(\bar{z})|} =
  \left(\frac{1}{2}\right)^{1+|S\setminus\operatorname{supp}(z)|}.
  \label{eq:offer-success}
\end{equation}
The thinning trial does not affect the path through the tree. If $z_S=\mathbf0$, the dynamic execution stops at this offer and records a clearance of the focal agent under the dynamic execution. Otherwise, the execution proceeds to the unique continuation child.

\end{enumerate}

If the execution reaches a leaf, it terminates without recording a dynamic clearance of the focal agent and without performing any further thinning trials. All thinning trials encountered during an execution are mutually independent. We use the conventions $z_\varnothing=\mathbf0$ and $\operatorname{supp}(z)\cap\varnothing=\varnothing$; consequently, an offer with $S=\varnothing$ always stops the dynamic execution and has thinning-success probability $p_\varnothing(z)=1/2$.
\end{definition}

The probability in \cref{eq:offer-success} has a direct interpretation. For the thinned offer to clear the healthy focal agent under the companion profile, $\bar{z}$, the thinning must retain the focal agent and discard every $j\in S$ for which $\bar{z}_j=1$ ($z_j = 0$). By contrast, every coordinate with $z_j=1$ becomes healthy under the companion profile, so its thinning outcome is irrelevant to clearance under the thinned offer. The focal agent contributes one required coin flip, and the $|S \cap \operatorname{supp}(\bar{z})| = |S\setminus\operatorname{supp}(z)|$ zero-coordinates of $z$ contribute one each.

Fix a reduced normalized tree $T$. For a profile $z$, let $I_T(z)\in\{0,1\}$ indicate whether the dynamic execution terminates at a stopping offer rather than at a leaf. Let $\mathcal O_T(z)$ be the collection of offer nodes encountered during the execution, including the stopping offer when one is reached. If $S_v$ denotes the label of offer node $v$, define
\begin{equation}
  F_T(z)
  :=
  \prod_{v\in\mathcal O_T(z)}
  \bigl(1-p_{S_v}(z)\bigr),
  \label{eq:failure-product}
\end{equation}
where we let the empty product be one. Because the thinning trials are independent, $F_T(z)$ is the probability that every encountered offer fails under the companion profile $\bar z$. Consequently, $1-F_T(z)$ is the probability that at least one thinning trial succeeds before the execution reaches a leaf or stops at a clearing offer. When the tree $T$ is fixed, we suppress the subscript and write $I(z)$ and $F(z)$. A fully worked three-coordinate example, including the execution path and the values of $I(z)$ and $F(z)$ for every profile, is given in \cref{app:worked-reduced-tree}.

\subsection{The Boolean-tree Charging Lemma}

With the definition of Reduced Normalized Trees in hand, we prove the key combinatorial lemma at the heart of our proof. In what follows, we say that a set $R \subseteq \{0,1\}^E$ is increasing if for $z \in R$ whenever $z' \geq z$ (i.e. $z'_i \geq z_i$ for all $i \in E$), then $z' \in R$. Similarly, $f: R \rightarrow [0,1]$ is decreasing if $z' \geq z$ implies $f(z') \leq f(z)$. 

\begin{lemma}[Monotone failure-weight invariant]
\label{lem:failure-weight}
Let $R\subseteq\{0,1\}^E$ be increasing and let $f:R\to[0,1]$ be decreasing. Put $M:=\max_{z\in R}f(z)$, with $M:=0$ if $R$ is empty. For every reduced normalized tree $T$,
\begin{equation}
  \sum_{z\in R}\left[f(z)F_T(z)(1+I_T(z))-1\right]
  \le -|R|(1-M).
  \label{eq:failure-invariant}
\end{equation}
In particular, on the full cube,
\begin{equation}
  \bigl|\{z:I_T(z)=1\}\bigr|
  \le 2\sum_{z\in\{0,1\}^E}\bigl(1-F_T(z)\bigr).
  \label{eq:half-thinning-count}
\end{equation}
\end{lemma}

\begin{proof}
For a current tree $T$, write
\[
  W_T(R,f):=\sum_{z\in R}\left[f(z)F_T(z)(1+I_T(z))-1\right]
\]
for the left side of \cref{eq:failure-invariant}, and below we abbreviate it to $W$. We prove the bound by induction on the remaining tree depth.

\smallskip
\noindent\emph{Base Case:}
The base case is a tree consisting of a single leaf in which case $I_T=0$ and $F_T=1$, so
\[
  \sum_{z\in R}(f(z)-1)\le |R|(M-1).
\]
\smallskip
\noindent\emph{Query Root:}
Let us now suppose first that the root is a query $A$.  Partition
\[
  R_0:=\{z\in R:z_A=\mathbf0\},
  \qquad
  R_+:=R\setminus R_0.
\]
Since $R$ is increasing, the set $R_+$ is also increasing. After fixing and deleting the coordinates $A$, the set $R_0$ is increasing on the \emph{residual cube} $\{0,1\}^{E\setminus A}$: its projection contains the remaining-coordinate vectors of profiles in $R_0$.  The zero-child normalization makes its continuation independent of the deleted coordinates. In addition, the restrictions (and, on $R_0$, projection) of $f$ remain decreasing.  If their maxima are $M_0$ and $M_+$, we use the fact that a query adds no thinning factor of its own and the induction hypothesis for the two children to prove the desired inequality:
\[
  W_T\le-|R_0|(1-M_0)-|R_+|(1-M_+)
    \le-|R|(1-M),
\]
\smallskip
\noindent\emph{Offer Root:}
Now suppose that the root is an offer $S$.  Write $k:=|S|$ and set
\[
  R_{\mathrm z}:=\{z\in R:z_S=\mathbf0\},
  \qquad R_{\mathrm p}:=R\setminus R_{\mathrm z},
  \qquad p_0:=2^{-k-1}.
\]
Profiles in $R_{\mathrm z}$ stop at this offer.  On $R_{\mathrm p}$ define
\[
  f'(z):=f(z)(1-p_S(z)).
\]
Let us start by considering the case where $R_{\mathrm p}$ is empty. One way this occurs is when $R$ is also empty. However, in this case, the lemma's desired inequality holds from the fact that $W_T(R,f) = 0$ (as the sum is empty), and $|R| = 0$. If $R_{\mathrm p}$ is empty  and $R$ is non-empty, then it must be the case that $S = \varnothing$. To see this, assume the contrary where $S \neq \varnothing$ and consider any $z\in R_{\mathrm z}$. Notice that flipping one coordinate of (the non-empty) $S$ (to 1) would produce a profile in $R$ (since the set is increasing), which in turn would necessarily be in $R_{\mathrm p}$, implying $R_{\mathrm p}$ is in fact non-empty, a contradiction. Thus let us continue with the case where $S=\varnothing$. 

When $S=\varnothing$, every $z \in R$ must accept at the root offer (so $I_T(z) = 1$ for all $z \in R$), and in addition the thinning clearance probability at this offer is given by $p_0=1/2$. As such we obtain
$$
W_T(R,f) = \sum_{z \in R}f(z)(1-p_0)(2) - 1
\leq |R|(M-1),
$$
which is the desired inequality of the lemma. Either way, we have shown the lemma holds when $R_{\mathrm p}$ is empty. 

Now let us continue to the case where $R_{\mathrm p}$ is nonempty. Once more, since $R$ is increasing, the set $R_{\mathrm p}$ is increasing.  In addition, as coordinates change from zero to one, fewer thinning coins are forced to discard companions, so $p_S$ is increasing and $1-p_S$ is decreasing.  Hence $f'=f\cdot(1-p_S)$ is decreasing.  Moreover, every $z\in R_{\mathrm p}$ has at least one one-coordinate in $S$, so $p_S(z)\ge2^{-k}$. In conjunction with the bound $f(z)\le M$, we get
\begin{equation}
  p_S(z)\ge2^{-k},
  \qquad M':=\max_{R_{\mathrm p}}f'
  \le M(1-2^{-k}).
  \label{eq:offer-max}
\end{equation}
To continue, partition the sum defining $W_T(R,f)$ according to
$R=R_{\mathrm z}\sqcup R_{\mathrm p}$:
\begin{align*}
  W_T(R,f)
  &=
  \sum_{z\in R}
  \left[
    f(z)F_T(z)\bigl(1+I_T(z)\bigr)-1
  \right]\\
  &=
  \sum_{z\in R_{\mathrm z}}
  \left[
    f(z)F_T(z)\bigl(1+I_T(z)\bigr)-1
  \right]\\
  &\qquad+
  \sum_{z\in R_{\mathrm p}}
  \left[
    f(z)F_T(z)\bigl(1+I_T(z)\bigr)-1
  \right].
\end{align*}

For the first sum, all $z \in R_{\mathrm z}$ clear the root offer, hence $I_T(z) = 1$. In addition, there is one thinning trial, which fails with probability $F(z) = 1-p_0$. It follows that the first sum is bounded by $|R_{\mathrm z}|\bigl[2M(1-p_0)-1\bigr]$. 

For continuing profiles, $z \in R_{\mathrm p}$, suppose that $T_{\mathrm p}$ is the tree of the continuation child. Given that an offer is made at the root, we have $F_T(z) = (1-p_S(z))F_{T_{\mathrm p}}(z)$, hence the second sum is:
\begin{align*}
  &\sum_{z\in R_{\mathrm p}}
  \left[
    f(z)F_T(z)\bigl(1+I_T(z)\bigr)-1
  \right]\\
  &\qquad=
  \sum_{z\in R_{\mathrm p}}
  \left[
    f(z)\bigl(1-p_S(z)\bigr)
    F_{T_{\mathrm p}}(z)
    \bigl(1+I_{T_{\mathrm p}}(z)\bigr)-1
  \right]\\
  &\qquad=
  \sum_{z\in R_{\mathrm p}}
  \left[
    f'(z)F_{T_{\mathrm p}}(z)
    \bigl(1+I_{T_{\mathrm p}}(z)\bigr)-1
  \right]\\
  &\qquad=
  W_{T_{\mathrm p}}(R_{\mathrm p},f').
\end{align*}
Thus a continuing profile on the other hand is exactly the child expression with weight $f'=f\cdot (1-p_S)$, which means we can directly apply the inductive hypothesis to bound it by $-|R_{\mathrm p}|(1-M')$. Putting everything together, we have the following bound:
\begin{equation}
  W_T\le |R_{\mathrm z}|\bigl[2M(1-p_0)-1\bigr]
        -|R_{\mathrm p}|(1-M').
  \label{eq:offer-induction}
\end{equation}

For every $z\in R_{\mathrm z}$, hold its coordinates in $E\setminus S$ fixed and replace its all-zero restriction on $S$ by any of the $2^k-1$ nonzero patterns. Upward closure places every such extension in $R_{\mathrm p}$. Extensions generated from distinct $z\in R_{\mathrm z}$ have different restrictions to $E\setminus S$, so the families are disjoint. Therefore
\begin{equation}
  |R_{\mathrm p}|\ge(2^k-1)|R_{\mathrm z}|.
  \label{eq:upward-count}
\end{equation}
Subtracting the target $-|R|(1-M)$ from the right side of \cref{eq:offer-induction}, and using \cref{eq:offer-max}, along with the fact that $|R| = |R_{\mathrm z}| + |R_{\mathrm p}|$, gives at most
\begin{align*}
 M|R_{\mathrm z}|(1-2p_0)-M|R_{\mathrm p}|2^{-k}
 &\le M|R_{\mathrm z}|\left[(1-2^{-k})-(2^k-1)2^{-k}\right]\\
 &=0.
\end{align*}
This proves the invariant, including the case $k=0$.

All that remains is to obtain \cref{eq:half-thinning-count}. Take $R=\{0,1\}^E$ and $f\equiv1$ in \cref{eq:failure-invariant}.  Its left side becomes
\[
  \sum_{z: I_T(z)=1}F(z)-\sum_z(1-F(z)),
\]
which the inequality proves is nonpositive.  With $H(z):=1-F(z)$, it follows that,
\[
  |\{z: I_T(z)=1\}|=\sum_{z: I_T(z)=1}(F(z)+H(z))
  \le\sum_zH(z)+\sum_{z: I_T(z)=1}H(z)
  \le2\sum_z H(z).
\]
Here the first inequality uses $\sum_{z: I_T(z)=1}F(z) \le\sum_z H(z)$, which is exactly the aggregate invariant; the second uses $\{z: I_T(z)=1\}\subseteq\{0,1\}^E$.
\end{proof}

\paragraph{Lemma intuition: failure-mass charging.}
The invariant we just proved (Equation \ref{eq:failure-invariant}) becomes more transparent after a simple rearrangement.  Define
\[
  g(z):=f(z)F_T(z).
\]
Although the lemma permits an arbitrary decreasing function $f$, in the induction one may regard $f(z)$ as the failure weight accumulated before entering the current subtree, as occurs in the offer-root induction. Since $F_T(z)$ is the probability that all thinning trials encountered within the subtree fail, $g(z)=f(z)F_T(z)$ is the total failure weight associated with profile $z$, combining the failure weight accumulated before the subtree with the failure probability contributed by the subtree itself.

Adding $|R|$ to both sides of the invariant and separating the profiles for which $I_T(z)=1$ gives the equivalent inequality
\begin{equation}
  \sum_{\substack{z\in R\\ I_T(z)=1}} f(z)F_T(z)
  \le
  \sum_{z\in R}\bigl(M-f(z)F_T(z)\bigr).
  \label{eq:failure-mass-charging}
\end{equation}
This expression provides a failure-mass charging interpretation of the lemma. Imagine that each profile begins with a budget of $M$ units. Every profile retains $f(z)F_T(z)$ units of failure mass, while a profile at which the dynamic execution clears the focal agent requires an additional copy of this mass. The right-hand side of \cref{eq:failure-mass-charging} is the total unused budget across all profiles, and the lemma says that this unused budget is large enough to pay for all of the additional failure mass carried by the dynamically stopping profiles.
Moreover, the charging is aggregate over $R$: we compare the total available and required mass rather than comparing the two summands profile by profile.

\subsection{From a Policy Tree to a Boolean Subcube}

A Boolean subcube of the full infection cube $\{0,1\}^{[n]}$ is specified by a set of free coordinates $E \subseteq [n]$ and a fixed assignment $a \in \{0,1\}^{[n]\setminus E}$. The corresponding subcube is
\[
C = C(a,E)
:=
\bigl\{
x \in \{0,1\}^{[n]} :
x_{[n]\setminus E} = a
\bigr\}.
\]
Thus, the coordinates in $E$ range freely over $\{0,1\}^E$, while the coordinates in $[n]\setminus E$ remain fixed at $a$.

For each free-coordinate assignment $z \in \{0,1\}^E$, let $x^a(z) \in C$ denote its unique extension to a full infection profile:
\[
x^a(z)_j
:=
\begin{cases}
z_j, & j \in E,\\
a_j, & j \notin E.
\end{cases}
\]
If $\bar z \in \{0,1\}^E$ is the coordinatewise complement of $z$, then the \emph{companion profile} to $x^a(z)$ within $C$ is
\[
\overline{x^a(z)}^{\,C}
:=
x^a(\bar z).
\]
In other words, the companion operation flips precisely the free coordinates and leaves the fixed coordinates unchanged. For the policy $\pi$ under consideration and a focal agent $i$, define, for every full infection profile $x \in \{0,1\}^{[n]}$,
\[
d_i(x)
:=
\1\bigl\{
\pi \text{ dynamically clears } i
\text{ under profile } x
\bigr\}.
\]
For any list of pools $P$ and any full infection profile
$y \in \{0,1\}^{[n]}$, let
\[
h_i(P,y)
\]
denote the probability, over the half-thinning coins, that the half-thinned version of $P$ clears agent $i$ when evaluated on profile $y$. In particular,
\[
h_i\bigl(P_\pi(x^a(z)),x^a(\bar z)\bigr)
\]
is the probability that the half-thinned path allocation selected under $x^a(z)$ clears $i$ when evaluated on its companion profile $x^a(\bar z)$.

\begin{lemma}[Subcube half-thinning]
\label{lem:subcube-half-thinning}
Let $\pi$ be a normalized deterministic policy, and let
\[
C = C(a,E)
:=
\bigl\{
x \in \{0,1\}^{[n]} :
x_{[n]\setminus E} = a
\bigr\}
\]
be a Boolean subcube with free-coordinate set $E$. For every focal agent $i$,
\begin{equation}
\sum_{z \in \{0,1\}^E}
d_i\bigl(x^a(z)\bigr)
\le
2
\sum_{z \in \{0,1\}^E}
h_i\bigl(P_\pi(x^a(z)),x^a(\bar z)\bigr).
\label{eq:subcube-half-thinning}
\end{equation}
\end{lemma}

\begin{proof}
We consider the three possible statuses of the focal coordinate: it may be fixed infected, fixed healthy, or free on the subcube.

\smallskip
\noindent\emph{Case 1: Focal Agent Fixed Infected.}
Suppose first that $i\notin E$ and $a_i=1$, so that $i$ is fixed infected throughout $C$. An infected agent cannot be cleared by a negative test, and hence
\[
  d_i\bigl(x^a(z)\bigr)=0
  \qquad
  \text{for every }z\in\{0,1\}^E.
\]
The left-hand side of \cref{eq:subcube-half-thinning} is therefore zero,
so the inequality is immediate.

\smallskip
\noindent\emph{Case 2: Focal Agent Fixed Healthy.}
Now suppose that $i\notin E$ and $a_i=0$, so that $i$ is fixed healthy. We reduce the part of the policy tree relevant to the clearance of $i$ to a reduced normalized tree on the free coordinates $E$. Along each policy path, we retain the execution only through the first negative pool containing $i$, if one is reached. After that pool, agent $i$ has already been cleared, so later tests are irrelevant to its dynamic clearance.

Consider first a tested pool not containing $i$. Fixed healthy members may be deleted. If the pool contains a fixed infected member, its outcome is always positive, so the corresponding node may be contracted to its positive child. Otherwise, after deleting the fixed coordinates, the pool becomes a query labelled by its free members $A\subseteq E$. Its outcome is negative exactly when $z_A=\mathbf 0$. Following such a negative outcome, every agent in $A$ has been cleared; normalization of $\pi$ therefore ensures that all later pool labels exclude $A$. This is precisely the zero-child exclusion required of a reduced normalized
tree.

Next consider a tested pool containing $i$. Fixed healthy companions may again be deleted. If the pool contains a fixed infected companion, it is always positive dynamically, so we contract the node to its positive child and omit its thinning trial. This leaves the dynamic clearance indicator unchanged. It can only decrease the static clearance probability, since thinning might otherwise discard the fixed infected companion, and therefore makes the desired inequality
harder to prove.

Every remaining pool containing $i$ becomes an offer labelled by its free companions $S\subseteq E$. Since $i$ is fixed healthy, the offer clears $i$ dynamically exactly when
\[
  z_S=\mathbf 0.
\]
Its paired evaluator is $x^a(\bar z)$. Under this profile, a free companion $j\in S$ is infected exactly when $z_j=0$. Consequently, the half-thinned offer clears $i$ exactly when thinning retains $i$ and discards every member of $S$ whose coordinate in $z$ is zero. This occurs with probability
\[
  2^{-1-|S\setminus\operatorname{supp}(z)|},
\]
as required by \cref{eq:offer-success} in the definition of a reduced normalized tree.

Let $T$ be the resulting reduced normalized tree. By construction,
\[
  I_T(z)=d_i\bigl(x^a(z)\bigr),
\]
while 
\[
  1-F_T(z)
  \le
  h_i\bigl(P_\pi(x^a(z)),x^a(\bar z)\bigr).
\]
Here the inequality arises because the reduction deletes some thinning opportunities and truncates the tree after the focal agent is dynamically cleared. Retaining those additional opportunities can only increase the static clearance probability. Applying \cref{eq:half-thinning-count} therefore gives
\begin{align*}
  \sum_{z\in\{0,1\}^E}d_i\bigl(x^a(z)\bigr)
  &=
  \sum_{z\in\{0,1\}^E}I_T(z)\\
  &\le
  2\sum_{z\in\{0,1\}^E}\bigl(1-F_T(z)\bigr)\\
  &\le
  2\sum_{z\in\{0,1\}^E}
  h_i\bigl(P_\pi(x^a(z)),x^a(\bar z)\bigr).
\end{align*}
This proves the lemma when the focal coordinate is fixed healthy.

\smallskip
\noindent\emph{Case 3: Focal Agent Free Coordinate.}
It remains to consider the case $i\in E$. Set
\[
  E':=E\setminus\{i\}.
\]
For every $v\in\{0,1\}^{E'}$, let $\bar v$ denote its coordinatewise complement in $\{0,1\}^{E'}$, and define two full profiles $x^H(v),x^I(v)\in C$ by
\[
  x^H(v)_i=0,
  \qquad
  x^I(v)_i=1,
\]
and
\[
  x^H(v)_{E'}=x^I(v)_{E'}=v,
  \qquad
  x^H(v)_{[n]\setminus E}
  =
  x^I(v)_{[n]\setminus E}
  =
  a.
\]
These profiles form the healthy and infected faces
\[
  C^H:=\{x^H(v):v\in\{0,1\}^{E'}\},
  \qquad
  C^I:=\{x^I(v):v\in\{0,1\}^{E'}\},
\]
and partition the original subcube:
\[
  C=C^H\sqcup C^I.
\]
The companion operation on $C$ exchanges the two faces:
\[
  \overline{x^H(v)}^{\,C}=x^I(\bar v),
  \qquad
  \overline{x^I(v)}^{\,C}=x^H(\bar v).
\]
Since the focal agent is infected under every profile in $C^I$,
\[
  d_i(x^I(v))=0.
\]
Similarly, when the selector is $x^H(v)$, its companion in the full subcube is $x^I(\bar v)$, under which the focal agent is infected. Therefore,
\[
  h_i\bigl(P_\pi(x^H(v)),x^I(\bar v)\bigr)=0.
\]
Consequently, the two sides of \cref{eq:subcube-half-thinning} reduce respectively to
\begin{align}
  \sum_{z\in\{0,1\}^E}d_i\bigl(x^a(z)\bigr)
  &=
  \sum_{v\in\{0,1\}^{E'}}d_i(x^H(v)),
  \label{eq:free-focal-left}\\
  \sum_{z\in\{0,1\}^E}
  h_i\bigl(P_\pi(x^a(z)),x^a(\bar z)\bigr)
  &=
  \sum_{v\in\{0,1\}^{E'}}
  h_i\bigl(P_\pi(x^I(v)),x^H(\bar v)\bigr).
  \label{eq:free-focal-right}
\end{align}

Regard $C^H$ as a subcube with free-coordinate set $E'$ and with the focal coordinate fixed healthy. The fixed-healthy case proved above then gives
\begin{equation}
  \sum_{v\in\{0,1\}^{E'}}d_i(x^H(v))
  \le
  2\sum_{v\in\{0,1\}^{E'}}
  h_i\bigl(P_\pi(x^H(v)),x^H(\bar v)\bigr).
  \label{eq:healthy-face-bound}
\end{equation}
We claim that, for every $v\in\{0,1\}^{E'}$,
\begin{equation}
  h_i\bigl(P_\pi(x^H(v)),x^H(\bar v)\bigr)
  \le
  h_i\bigl(P_\pi(x^I(v)),x^H(\bar v)\bigr).
  \label{eq:free-focal-path-domination}
\end{equation}

To prove the claim, compare the executions of $\pi$ under $x^H(v)$ and $x^I(v)$. These profiles differ only in the infection status of $i$. Let $t$ be the first pool that clears $i$ in the execution under $x^H(v)$, if such a pool exists. Before $t$, a tested pool not containing $i$ has the same outcome under both profiles. A tested pool containing $i$ must be positive under $x^H(v)$, since otherwise it would clear $i$ before $t$; it is also positive under $x^I(v)$ because $i$ is infected there. Inductively, the two executions therefore select exactly the same pools through and including $t$. At $t$, the outcome is negative under $x^H(v)$ and positive under $x^I(v)$.

Evaluate both selected path allocations on the same profile $x^H(\bar v)$, and couple their half-thinning coins on their common pool occurrences. Because $\pi$ is normalized, no pool selected after $t$ under $x^H(v)$ contains $i$. Hence every thinning opportunity that can clear $i$ in $P_\pi(x^H(v))$ occurs in the common prefix through $t$, which is also contained in $P_\pi(x^I(v))$. Thus every thinning outcome that clears $i$ using the former path also clears it using the latter path. The infected execution may provide additional opportunities after $t$, proving \cref{eq:free-focal-path-domination}.

If no such pool $t$ exists, every pool containing $i$ tests positive under both profiles, while pools excluding $i$ again have identical outcomes. The two executions then follow the same path for their entire duration, and \cref{eq:free-focal-path-domination} holds with equality.

Summing \cref{eq:free-focal-path-domination} over $v$ and combining it with \cref{eq:healthy-face-bound} gives
\[
  \sum_{v\in\{0,1\}^{E'}}d_i(x^H(v))
  \le
  2\sum_{v\in\{0,1\}^{E'}}
  h_i\bigl(P_\pi(x^I(v)),x^H(\bar v)\bigr).
\]
Together with \cref{eq:free-focal-left,eq:free-focal-right}, this is exactly
\cref{eq:subcube-half-thinning}.
\end{proof}

\subsection{Symmetrization for Arbitrary Product Priors}

The subcube lemma is an unweighted statement: it compares uniform sums over a Boolean subcube. We now show that two independent draws from an arbitrary heterogeneous product prior can be decomposed into a random subcube together with a uniform orientation of its free coordinates. This converts the subcube inequality into the desired product-distribution inequality without imposing any bounds on individual infection
probabilities.

\begin{theorem}[Per-agent half-thinning inequality]
\label{thm:per-agent-half-thinning}
Let $Z$ and $Z'$ be independent infection profiles drawn from the instance's product prior. For every normalized deterministic policy $\pi$ and every agent $i$,
\begin{equation}
  \Prob\bigl(\pi\text{ dynamically clears }i\bigr)
  \le
  2\E\left[h_i\bigl(P_\pi(Z),Z'\bigr)\right].
  \label{eq:per-agent-half-thinning}
\end{equation}
\end{theorem}

\begin{proof}
For each coordinate $j\in[n]$, define the unordered pair
\[
  U_j
  :=
  \bigl(
    \min\{Z_j,Z'_j\},
    \max\{Z_j,Z'_j\}
  \bigr)
  \in\{(0,0),(0,1),(1,1)\},
\]
and let $U:=(U_1,\ldots,U_n)$. We condition on a realization of $U$ having positive probability.

This realization determines a set of free coordinates
\[
  E:=\{j\in[n]:U_j=(0,1)\},
\]
namely the coordinates on which $Z$ and $Z'$ disagree. On every coordinate $j\notin E$, the two profiles agree, and we define their common fixed value by
\[
  a_j:=Z_j=Z'_j.
\]
Thus $U$ determines the Boolean subcube
\[
  C(a,E)
  =
  \bigl\{x\in\{0,1\}^{[n]}:
  x_{[n]\setminus E}=a\bigr\}.
\]

It remains to determine the conditional distribution of the free coordinates. For every $j\in E$, the two possible orientations have the same probability:
\[
  \Prob(Z_j=0,Z'_j=1)
  =
  \Prob(Z_j=1,Z'_j=0),
\]
because $Z_j$ and $Z'_j$ are independent draws from the same marginal distribution. Consequently,
\[
  \Prob(Z_j=0\mid U_j=(0,1))
  =
  \Prob(Z_j=1\mid U_j=(0,1))
  =
  \frac12.
\]
Moreover, the coordinate pairs $(Z_j,Z'_j)$ are independent across $j$. Conditioning separately on their unordered values therefore leaves these orientation choices independent across the coordinates
in $E$.

Let
\[
  V:=Z_E\in\{0,1\}^E.
\]
Conditional on $U$, the vector $V$ is uniform on $\{0,1\}^E$, and the two full profiles can be written as
\[
  Z=x^a(V),
  \qquad
  Z'=x^a(\bar V).
\]
Thus, conditional on $U$, the first profile is uniform on the subcube $C(a,E)$ and the second is exactly its companion within that subcube. Applying \cref{lem:subcube-half-thinning} to $C(a,E)$ and dividing its two sums by $2^{|E|}$ gives
\begin{align*}
  \E\left[d_i(Z)\mid U\right]
  &=
  \frac{1}{2^{|E|}}
  \sum_{v\in\{0,1\}^E}
  d_i\bigl(x^a(v)\bigr)\\
  &\le
  \frac{2}{2^{|E|}}
  \sum_{v\in\{0,1\}^E}
  h_i\bigl(P_\pi(x^a(v)),x^a(\bar v)\bigr)\\
  &=
  2\E\left[
    h_i\bigl(P_\pi(Z),Z'\bigr)
    \mathrel{\Big|}U
  \right].
\end{align*}
Finally, taking expectations over $U$ and applying the law of iterated
expectation yields
\begin{align*}
  \Prob\bigl(\pi\text{ dynamically clears }i\bigr)
  &=
  \E[d_i(Z)]\\
  &\le
  2\E\left[h_i\bigl(P_\pi(Z),Z'\bigr)\right],
\end{align*}
which proves \cref{eq:per-agent-half-thinning}.
\end{proof}

\begin{theorem}[Uniform factor-two adaptivity gap]
\label{thm:factor-two}
For every population $J$, test budget $B$, and pool-size cap $G$,
\begin{equation}
  \OPTdyn(B,J)\le2\OPTstar(B,J),
  \qquad\text{equivalently}\qquad
  \Agap(B,J)\le2.
  \label{eq:factor-two}
\end{equation}
\end{theorem}

\begin{proof}
Fix an arbitrary deterministic policy $\pi$. By
\cref{lem:normalize-policy}, we may assume without loss of generality that $\pi$ is normalized. Sample the virtual profile $Z$, follow $P_\pi(Z)$, and half-thin every incidence. The resulting random list contains at most $B$ feasible pools and is sampled independently of the true evaluation profile $Z'$. It is therefore a randomized feasible static allocation. More explicitly, after fixing $Z$ and all thinning coins, one obtains a deterministic list of at most $B$ pools, each of size at most $G$, whose expected welfare under $Z'$ is at most $\OPTstar(B,J)$.  Averaging this pointwise bound over $Z$ and the thinning coins preserves the inequality.

Multiply \cref{eq:per-agent-half-thinning} by $u_i$ and sum over agents:
\[
  V(\pi)
  \le2\E\left[\sum_i u_i h_i(P_\pi(Z),Z')\right]
  \le2\OPTstar(B,J).
\]
Taking $\pi$ to be optimal proves the theorem.
\end{proof}

\begin{corollary}[Greedy policies versus the dynamic optimum]
\label{cor:greedy-dynamic}
The exact-oracle dynamic and static greedy policies satisfy
\[
  V(\Greedy)\ge\frac1{2(e+1)}\OPTdyn(B,J),
  \qquad
  V(\StaticGreedy)\ge\frac1{2(e+1)}\OPTdyn(B,J).
\]
Equivalently, each is a $2(e+1)$-approximation to the optimal dynamic policy.
\end{corollary}

\begin{proof}
For $\Greedy$, combine \cref{thm:main-greedy,thm:factor-two}. For $\StaticGreedy$, combine \cref{eq:static-greedy-finster,thm:factor-two}.
\end{proof}

\section{Exact Small-instance Experiments}
\label{sec:experiments}

The approximation theorems are worst-case guarantees.  To see how the relevant benchmarks and policies compare away from specially constructed instances, we perform a reproducible experiment in the regime where every expectation and posterior can be computed exactly.

\subsection{Design}

For each instance, health probabilities are drawn independently and uniformly from $[0.05,0.95]$, and utilities are drawn independently from the log-uniform distribution on $[0.1,2]$.  We use the fixed seed $20260830$ and three settings: 300 instances with $(n,B,G)=(4,2,4)$, 200 with $(5,3,5)$, and 100 with $(6,3,3)$.  These are random interior instances, not an adversarial search.

The posterior-state dynamic program described in \cref{app:exact-dp} enumerates all $2^n$ health profiles and returns $\OPTdyn$.  Exhaustive static search returns both the overlapping benchmark $\OPTstar$ and the non-overlapping benchmark $\OPTno$.  We compare these with three policies: $\Greedy$; $\MPGreedy$, evaluated here using \emph{exact} posterior marginals; and $\StaticGreedy$ from \cref{subsec:greedy-comparison}, implemented with an exact single-test oracle.  Thus the experiment introduces no Monte Carlo error.  Table \ref{tab:exact-small} reports each method's welfare divided instance by instance by $\OPTdyn$, then averaged; parentheses contain standard errors across instances.

\begin{table}[ht]
  \centering
  \footnotesize
  \setlength{\tabcolsep}{3pt}
  \caption{Exact small-instance performance.  Entries in the five policy columns are mean
  $V/\OPTdyn$ (standard error).  The final column is the largest observed
  $\OPTdyn/\OPTstar$ in that setting.}
  \label{tab:exact-small}
  \begin{tabular}{lrrrrrr}
    \toprule
    $(n,B,G)$ & $\OPTstar$ & $\OPTno$ & Static greedy & $\Greedy$ & $\MPGreedy$ & Max gap \\
    \midrule
    $(4,2,4)$ & .9948 (.0007) & .9932 (.0009) & .9529 (.0043) & .9881 (.0015) & .9880 (.0015) & 1.0744 \\
    $(5,3,5)$ & .9908 (.0009) & .9895 (.0010) & .9242 (.0069) & .9879 (.0014) & .9876 (.0014) & 1.0501 \\
    $(6,3,3)$ & .9894 (.0013) & .9838 (.0019) & .9126 (.0090) & .9863 (.0019) & .9858 (.0020) & 1.0507 \\
    \bottomrule
  \end{tabular}
\end{table}

\subsection{Results and Interpretation}

All methods perform much better on these draws than their worst-case guarantees alone would predict. The minimum observed fraction of $\OPTdyn$ achieved by $\Greedy$ was $0.8343$, $0.8977$, and $0.8962$ in the three settings, respectively; the theorem guarantees only $1/[2(e+1)]\approx0.1345$.  The same worst-case fraction applies to $\StaticGreedy$ by \cref{cor:greedy-dynamic}.  Likewise, the largest observed adaptivity gap was $1.0744$, far below the universal upper bound of two.  These observations describe the sampled distributions and are not evidence for a stronger worst-case theorem.

The experiment also confirms that there is no simple empirical dominance between dynamic greedy and the optimal static overlapping allocation.  In the three settings, $\Greedy$ was strictly better than $\OPTstar$ on $16.7\%$, $27.0\%$, and $33.0\%$ of instances, while $\OPTstar$ was strictly better on $21.3\%$, $26.0\%$, and $37.0\%$; the remaining instances were numerical ties. Exact-Joint and Marginal-Product Greedy were nearly indistinguishable on average, but not pointwise: $\Greedy$ was strictly better on 2, 11, and 8 instances, respectively, while $\MPGreedy$ was strictly better on one instance in the $(5,3,5)$ setting.  The equality of their first action under the independent prior does not persist after positive tests create posterior dependence.

The direct comparison between $\Greedy$ and $\StaticGreedy$ appears in \cref{tab:greedy-head-to-head}.  Exact-Joint Greedy was weakly better on all 600 sampled instances, but a finite random experiment cannot establish pointwise dominance; the general comparison remains open.

\subsection{Scalable Empirical Directions}

The exact comparison deliberately separates the marginal-product scoring rule from posterior estimation error.  For larger populations, a practical version estimates posterior marginals by Gibbs sampling and repeatedly applies a single-test optimizer to their product.  We call this \emph{Gibbs-Marginal Greedy}.  It is an empirical heuristic: neither replacing a joint probability by a product of marginals nor estimating those marginals by sampling preserves the hypotheses of \cref{thm:main-greedy}. In addition, a second empirical direction uses Proximal Policy Optimization (PPO) with bucketed agent features and sequential pool construction.  This reduces the state and action spaces enough to train on larger populations, but it discards agent-level information and introduces training and representation choices absent from the exact comparison.  
The exact dynamic program, Gibbs implementation strategy, PPO formulation, and reproducibility details are collected in \cref{app:computation}.  

\section{Conclusion}
\label{sec:conclusion}

Dynamic welfare-maximizing pooled testing admits a clean universal comparison despite the posterior dependence created by adaptive, overlapping tests.  The optimal dynamic policy is never more than twice as valuable as the optimal static overlapping allocation.  Adaptivity can provide a strict benefit, as the three-agent example demonstrates, but that benefit is uniformly bounded across all populations, test budgets, and pool-size caps.

The surrounding structural results help explain this conclusion.  Posterior domination limits the immediate welfare available from any adaptive test, while the path-disjoint equivalence shows that a strict adaptive advantage requires re-pooling agents after positive outcomes.  Exact-Joint Greedy turns these structural properties into a constant-factor guarantee, although the computation of joint posterior pool probabilities remains its principal practical obstacle.  The factor-two theorem also transfers the existing static Greedy guarantee into a $2(e+1)$-approximation of the optimal dynamic policy.  In the exact small-instance study, the greedy policies and static benchmarks were generally close to the dynamic optimum, although these finite experiments cannot establish a universal pointwise ordering among the algorithms.

The principal theoretical open problem is to determine whether the factor-two adaptivity bound is tight or instead reflects a loss specific to the half-thinning comparison.  Other directions include developing efficient approximations to joint conditional gains that preserve the subset stability required by the greedy analysis, and resolving the pointwise relationship between $\Greedy$ and $\StaticGreedy$.  

\begingroup
\footnotesize
\bibliographystyle{plainnat}
\bibliography{references}
\endgroup

\appendix
\section{Supporting Proofs and Supplementary Theory}
\label{app:technical-proofs}

This appendix supplies supporting proofs and extensions for the structural and greedy results in the main text.  It provides boundary regimes, complexity results, and approximate-oracle qualifications. 

\subsection{Boundary regimes}
\label{sec:boundaries}

The two bounds proved in \cref{thm:main-master-bounds} give the refinements collected in \cref{thm:main-boundaries}.  We prove all four parts here.  We also use the following equality observation later: equality in $V(\pi)\le Bu^*$ forces every selected uncleared pool to have prior value $u^*$ at every step.

\begin{proof}[Proof of \cref{thm:main-boundaries}]
For part (i), \cref{thm:main-master-bounds} gives $\OPTdyn\le T_{B\Lambda}\le T_B$.  Individually testing the agents with the $\min\{B,n\}$ largest values of $m_i=u_iq_i$ attains $T_B$, so all three nested benchmarks equal $T_B$.  If every $q_i\le1/2$, then every pool $A$ of size $k$ satisfies
\[
  \sum_{i\in A}q_{A\setminus\{i\}}
  \le k2^{-(k-1)}\le1,
\]
and hence $\Lambda\le1$.

For part (ii), when $B=1$ all three policy classes choose a single pool, so their optima coincide. When $B\ge n$, testing every agent individually attains $\sum_i u_iq_i$.  No policy can do better, because an agent contributes utility only if that agent is healthy.  Thus the adaptivity gap is one at both budget endpoints.

For part (iii), the assumed disjoint pools form a feasible non-overlapping allocation with additive welfare at least $(1-\gamma)Bu^*$.  The per-test bound in \cref{thm:main-master-bounds} gives $\OPTdyn\le Bu^*$, so
\[
  \OPTno(B,J)\ge(1-\gamma)Bu^*
  \ge(1-\gamma)\OPTdyn(B,J).
\]
Therefore $\Ano\le1/(1-\gamma)$, and $\Agap\le\Ano$ because $\OPTstar\ge\OPTno$.  When $\gamma=0$, the lower and upper bounds meet at $Bu^*$.

For part (iv), first suppose $\OPTno(B,J)=0$.  By the ratio convention in \cref{sec:model}, $\Agap=\Ano=1$, while every expression on the right of \cref{eq:main-high-q} is at least one, so the conclusion is immediate.  We may therefore assume $\OPTno(B,J)>0$.  Fix $s\in\{1,\ldots,G\}$.  The population-size assumption permits us to partition the $Bs$ agents with largest $m_i$ into $B$ disjoint pools of size $s$.  The welfare of these pools is at least
\[
  \theta^{s-1}T_{Bs},
\]
because every companion product $q_{t\setminus\{i\}}$ is at least $\theta^{s-1}$.  Therefore $\OPTno(B,J)\ge\theta^{s-1}T_{Bs}$.  By \cref{thm:main-master-bounds} and \cref{eq:top-sum-comparison},
\[
  \Ano(B,J)
  =\frac{\OPTdyn(B,J)}{\OPTno(B,J)}
  \le\frac{T_{B\Lambda}}{\theta^{s-1}T_{Bs}}
  \le\frac{\max\{1,\Lambda/s\}}{\theta^{s-1}}.
\]
Minimizing over $s$ proves the middle bound in \cref{eq:main-high-q}.  Since $\Lambda\le G$ and $s\le G$, we also have $\max\{1,\Lambda/s\}\le G/s$.  Finally, $\Agap\le\Ano$ and \cref{thm:factor-two} gives $\Agap\le2$, proving the stated minimum with the universal bound.
\end{proof}

\begin{remark}
Replication in part (iii) is a structural condition, not an automatic consequence of a large population.  Large populations may make near-copies likely in a specified sampling model, but the theorem requires the copies themselves.
\end{remark}

\subsection{Computational complexity}
\label{sec:complexity}

The dynamic problem inherits static hardness and also admits a direct exact-threshold reduction.

\begin{theorem}[NP-hardness]
\label{thm:np-hard}
For every fixed $G\ge3$, deciding whether $\OPTdyn(B,J)\ge\tau$ is NP-hard.
\end{theorem}

\begin{proof}[Proof sketch]
Use the strongly NP-hard \textsc{3-Partition} construction from \cite{finster2026static}.  Given numbers $a_1,\ldots,a_{3m}$ and target $T$, choose common health parameters so that the prior value of a pool with utility sum $x$ is $f(x)=xr^x$, uniquely maximized at the integer $x=T$.  Set $B=m$ and threshold $mf(T)$.

A yes-instance has $m$ disjoint triples of value $f(T)$, so a static policy reaches the threshold. Conversely, the per-test part of \cref{thm:main-master-bounds} gives $\OPTdyn\le mf(T)$.  Equality forces every uncleared pool $A_b$ chosen at every positive-probability history to have utility sum $T$.  Follow the all-negative path, which has positive probability.  The sets $A_1,\ldots,A_m$ are disjoint by definition because negative tests clear their members; each is a triple summing to $T$.  They form a valid 3-partition.
\end{proof}

\begin{proposition}[Evaluation hardness]
For $G\ge3$, evaluating a given dynamic policy is \#P-hard, and its threshold problem is PP-hard. Computing the optimal first test is W[1]-hard in $G$.
\end{proposition}

\begin{proof}
A static overlapping allocation is a dynamic policy that ignores all observations.  The claims therefore follow from the corresponding static results in \cite{finster2026static}; the first-test claim follows by setting $B=1$.
\end{proof}

\subsection{Approximate oracles and sampling error}
\label{sec:approx-oracle}

The complete exact-oracle argument appears in the main text as the proof of
\cref{thm:main-greedy}.  We now show how that argument changes for an approximate oracle and why sampled estimates require additional care.

\begin{hypothesis}[Subset-closed approximate joint oracle]
\label{hyp:approx-oracle}
At every history $H$, the oracle returns a feasible pool $t$ such that
\begin{enumerate}[label=(\roman*)]
\item $g(t\mid H)\ge(1-\varepsilon)\max_{|s|\le G}g(s\mid H)$;
\item $g(t\mid H)\ge g(s\mid H)$ for every $s\subseteq t$.
\end{enumerate}
\end{hypothesis}

\begin{proposition}[Conditional approximate-oracle guarantee]
\label{prop:approx-oracle}
If every greedy step satisfies Hypothesis~\ref{hyp:approx-oracle}, then
\[
  V\ge\frac{1}{e+(1-\varepsilon)^{-1}}\OPTstar(B,J).
\]
\end{proposition}

\begin{proof}
Condition (ii) permits the proof of \cref{lem:main-split-stability} (split stability) to be repeated for the returned pool, yielding the same risk bound as \cref{eq:main-split-stability}.  Because every $S\subseteq D_j$ is also a subset of the returned pool, it also preserves the comparison $u(S)\le g_j$ used in the auxiliary-subset argument leading to \cref{eq:main-fresh-charge}. Thus the fresh-agent loss remains at most $e g_j$. Condition (i) bounds the deleted-pool loss by $g_j/(1-\varepsilon)$.  Consequently
\[
  F_j-F_{j+1}\le\left(e+\frac1{1-\varepsilon}\right)g_j,
\]
and telescoping proves the claim.
\end{proof}

\subsection{Exoneration: posterior infection probabilities are not generally monotonic}
\label{sec:exoneration}

Posterior domination bounds every history by the priors, but posteriors are not monotone along a history.  

\begin{proposition}[Exoneration]
\label{prop:exoneration}
There are histories $H\subset H'$ and an uncleared agent $a$ for which
\[
  \Prob(X_a=1\mid H')>\Prob(X_a=1\mid H).
\]
\end{proposition}

\begin{proof}
Let $q_a=q_x=1/2$ and let $H$ consist of the positive outcome of pool $\{a,x\}$.  Then
\[
  \Prob(X_a=1\mid H)=\frac{q_a(1-q_x)}{1-q_aq_x}=\frac13.
\]
Append the positive singleton test $\{x\}$.  Agent $x$ is then known infected, fully explaining the first positive pool, and $\Prob(X_a=1\mid H')=q_a=1/2$.
\end{proof}

\section{Reduced Normalized Tree Example}
\label{app:worked-reduced-tree}

The following example illustrates separately how a profile determines the path through a reduced normalized tree, whether the dynamic experiment reaches a stopping offer, and the probability that at least one thinning trial succeeds under the companion profile.

\begin{example}
Let $E=\{a,b,c\}$. Write $Q_A$ for a query with label $A$ and $O_S$ for an offer with label $S$. Consider the following reduced normalized tree:

\begin{enumerate}[label=(\arabic*),leftmargin=*]
\item The root is the query $Q_{\{a\}}$.
\item Its zero child is the offer $O_{\{b\}}$. The continuation child
of this offer is $O_{\{c\}}$, whose continuation child is a leaf.
\item The positive child of the root is the offer $O_{\{b,c\}}$, whose
continuation child is a leaf.
\end{enumerate}

Thus, when $z_a=0$, the execution first reaches $O_{\{b\}}$. It stops there if $z_b=0$; otherwise, it continues to $O_{\{c\}}$ and stops there if $z_c=0$. When $z_a=1$, the execution instead reaches $O_{\{b,c\}}$, where it stops if $z_b=z_c=0$. The tree is normalized: after the zero outcome at $Q_{\{a\}}$, every subsequent label is disjoint from $\{a\}$.

For each profile $z$, the following table records the offer nodes encountered, their thinning-success probabilities in the order encountered, the dynamic stopping indicator $I(z)$, and the failure probability
\[
  F(z)=\prod_{S\text{ encountered}}\bigl(1-p_S(z)\bigr).
\]

\begin{center}
\small
\begin{tabular}{ccccc}
\toprule
$z=(z_a,z_b,z_c)$
  & Offers encountered
  & Thinning success probabilities
  & $I(z)$
  & $F(z)$ \\
\midrule
$000$ & $\{b\}$             & $1/4$       & $1$ & $3/4$ \\
$001$ & $\{b\}$             & $1/4$       & $1$ & $3/4$ \\
$010$ & $\{b\},\{c\}$       & $1/2,1/4$   & $1$ & $3/8$ \\
$011$ & $\{b\},\{c\}$       & $1/2,1/2$   & $0$ & $1/4$ \\
$100$ & $\{b,c\}$           & $1/8$       & $1$ & $7/8$ \\
$101$ & $\{b,c\}$           & $1/4$       & $0$ & $3/4$ \\
$110$ & $\{b,c\}$           & $1/4$       & $0$ & $3/4$ \\
$111$ & $\{b,c\}$           & $1/2$       & $0$ & $1/2$ \\
\bottomrule
\end{tabular}
\end{center}

For example, consider $z=010$. The root query takes its zero child because $z_a=0$. At the first offer, $z_b=1$, so the dynamic experiment does not stop. Since $b$ is healthy in the companion profile $\bar z=101$, the thinning trial only needs to retain the focal agent and therefore succeeds with probability $1/2$. The execution then reaches $O_{\{c\}}$, where it stops because $z_c=0$. Now $c$ is infected under $\bar z$, so thinning must retain the focal agent and discard $c$, giving success probability $1/4$. Hence
\[
  F(010)
  =\left(1-\frac12\right)\left(1-\frac14\right)
  =\frac38.
\]

As a second example, consider $z=100$. The root query takes its positive child, and $O_{\{b,c\}}$ is a stopping offer because $z_b=z_c=0$. Both $b$ and $c$ are infected under the companion profile $\bar z=011$. A thinning success therefore requires retaining the focal agent while discarding both companions, which occurs with probability
\[
  p_{\{b,c\}}(100)=\left(\frac12\right)^3=\frac18.
\]
Consequently, $I(100)=1$ and $F(100)=7/8$.

The profiles on which the dynamic experiment reaches a stopping offer are
\[
  \{z:I(z)=1\}=\{000,001,010,100\},
\]
so
\[
  \bigl|\{z:I(z)=1\}\bigr|=4.
\]
Meanwhile,
\begin{align*}
  \sum_{z\in\{0,1\}^E}\bigl(1-F(z)\bigr)
  &=
  \frac14+\frac14+\frac58+\frac34
  +\frac18+\frac14+\frac14+\frac12\\
  &=3.
\end{align*}
The factor-two counting conclusion of
\cref{lem:failure-weight} therefore reads
\[
  \bigl|\{z:I(z)=1\}\bigr|
  =4
  \le
  2\sum_{z\in\{0,1\}^E}\bigl(1-F(z)\bigr)
  =6.
\]
\end{example}

\section{Computational methods and implementation details}
\label{app:computation}

\subsection{Exact posterior-state dynamic programming}
\label{app:exact-dp}

The experiment in \cref{sec:experiments} uses exact small-instance optimization to recover an optimal dynamic policy, evaluate every candidate policy under the true posterior, and compute the static benchmarks without sampling.

A history $H$ determines the set $\Omega(H)\subseteq\{0,1\}^n$ of health profiles consistent with the observed outcomes and a set $C$ of agents already cleared.  Product-prior probabilities are retained as weights on $\Omega(H)$.  For a candidate pool $t$, let
\[
  \Omega_-(H,t)=\{x\in\Omega(H):x_i=1\text{ for every }i\in t\},
  \qquad
  \Omega_+(H,t)=\Omega(H)\setminus\Omega_-(H,t),
\]
and let $p_-(H,t)$ be the posterior probability of $\Omega_-(H,t)$.  With $b$ tests remaining, backward induction gives
\begin{align}
V_b(\Omega,C)=\max_{\substack{t\subseteq[n]\setminus C\\1\le |t|\le G}}
\bigl\{&p_-(H,t)\bigl[\textstyle\sum_{i\in t}u_i+
V_{b-1}(\Omega_-,C\cup t)\bigr] \notag\\
&+(1-p_-(H,t))V_{b-1}(\Omega_+,C)\bigr\},
\label{eq:exact-dp}
\end{align}
with $V_0=0$.  Memoizing $(\Omega,C,b)$ and storing a maximizing pool at each state returns an explicit optimal policy tree, not only its value.  Replacing the maximizing action by the pool with largest true immediate conditional gain evaluates $\Greedy$ on the same state space.

This procedure enumerates all $2^n$ profiles and all feasible pools, so it does not result in an algorithm which scale well with population sizes. We use this algorithm for benchmarking when populations are small.  A dependency-free implementation is included with the source.

\subsection{Gibbs-Marginal Greedy}
\label{app:gibbs-greedy}

Exact joint-posterior optimization is exponential in the population size.  A scalable heuristic instead scores a pool by
\[
  \left(\prod_{i\in U(t,H)}\Prob(X_i=1\mid H)\right)
  \sum_{i\in U(t,H)}u_i,
\]
where the posterior marginals are estimated after every observed outcome.  The Gibbs state is a health assignment consistent with all positive pools after cleared agents have been removed.  When updating one unresolved agent, the sampler forces infection if changing that coordinate to healthy would leave a positive pool with no infected member; otherwise it resamples from the prior Bernoulli law.  Empirical healthy frequencies estimate the marginals.  A single-test optimizer such as that of \citet{welfaremax} then chooses the next pool, and the process repeats online.

This implementation is more scalable than exact posterior-state optimization, but it has two distinct approximation layers: the product-of-marginals objective and Monte Carlo posterior estimation.  Its performance can also depend on convergence diagnostics, the one-step pool optimizer, and tie breaking.  We therefore refer to it as Gibbs-Marginal Greedy and treat it as an empirical method rather than an implementation of the theorem for $\Greedy$.

\subsection{PPO policies}
\label{app:ppo}

The PPO formulation constructs one pool sequentially from buckets of agents and uses separate policies indexed by the remaining test budget.  Bucket counts compress the state, while sequential pool construction replaces selection from all feasible subsets by a sequence of smaller actions. Training instances sample latent health profiles, and the reward is the realized welfare of the complete downstream policy rather than only the current test's reward.

This design reaches populations beyond the exact solver, but its abstraction discards agent-level information and uses posterior-marginal products.  Results depend on the number of buckets, network architecture, training horizon, random seeds, and compute budget.  PPO is consequently a flexible empirical baseline.

\subsection{Archived exploratory results from the earlier pipeline}
\label{app:archived-experiments}

For completeness, we retain a selected record of the computational work reported in earlier drafts.  These numbers answer different questions from the exact experiment in \cref{sec:experiments}.  In particular, the earlier ``Greedy Dynamic'' implementation multiplied Gibbs-estimated posterior marginals and is therefore what we now call Gibbs-Marginal Greedy; it was not an implementation of $\Greedy$.  The instance distributions also differ, and the large-instance study reports realized rather than exactly integrated welfare.  The tables below should therefore be read within each experimental design, not compared numerically with \cref{tab:exact-small}.

The retained large-instance study drew 10,000 populations with $n=50$, $B=5$, utilities uniform on $\{1,2,3\}$, and health probabilities uniform on $[0,1]$.  PPO~4 and PPO~5 used four and five health-probability buckets, respectively; the ``$+$'' variants were trained for 50 million rather than 20 million steps.  The archived implementation did not reuse the same realized health vector when evaluating Gibbs-Marginal Greedy and the static MILP on a given population.  Thus \cref{tab:archived-large} supports comparisons of aggregate means under the common population distribution, but not paired differences, per-instance win rates, or scatterplots.  Gibbs-Marginal Greedy has the largest reported aggregate mean in both settings.

\begin{table}[ht]
  \centering
  \small
  \caption{Earlier exploratory large-instance results.  Entries are average realized welfare over
  10,000 sampled instances and health realizations, with standard errors in parentheses.}
  \label{tab:archived-large}
  \begin{tabular}{llr}
    \toprule
    Setting & Method & Realized welfare \\
    \midrule
    $(n,B,G)=(50,5,5)$
      & Static MILP benchmark & 20.58 (0.0799) \\
      & Gibbs-Marginal Greedy & 21.15 (0.0892) \\
      & PPO 4 & 18.85 (0.0748) \\
      & PPO 4+ & 18.95 (0.0725) \\
      & PPO 5 & 18.60 (0.0706) \\
      & PPO 5+ & 19.12 (0.0721) \\
    \addlinespace
    $(n,B,G)=(50,5,3)$
      & Static MILP benchmark & 20.26 (0.0734) \\
      & Gibbs-Marginal Greedy & 20.52 (0.0742) \\
      & PPO 4 & 16.92 (0.0523) \\
      & PPO 4+ & 16.92 (0.0536) \\
      & PPO 5 & 17.28 (0.0545) \\
      & PPO 5+ & 17.30 (0.0540) \\
    \bottomrule
  \end{tabular}
\end{table}

\subsection{Evaluation and reproducibility considerations}

Code for the exact solver and computational approaches can be found at \url{https://github.com/nrlopez03/pooled-testing}.  The exact solver has no commercial-solver dependency, and the source bundle also contains the fixed-seed raw and summary CSV files underlying \cref{tab:exact-small}, together with the aggregate CSV underlying \cref{tab:archived-large}.
\end{document}